\documentclass[pra,twocolumn,10pt,aps,longbibliography]{revtex4-1}
\usepackage{graphicx}
\usepackage{amsmath} 
\usepackage{amssymb}
\usepackage{amsfonts}
\usepackage{color}
\begin{document}
\title{Dynamics of defect-induced dark solitons in an exciton-polariton condensate}
\author{Andrzej Opala$^{1,2}$, Maciej Pieczarka$^1$, Nataliya Bobrovska$^2$, Micha{\l} Matuszewski$^{2}$\footnote{corresponding author, email: opala@ifpan.edu.pl}}
\affiliation{
$^1$Laboratory for Optical Spectroscopy of Nanostructures, Department of Experimental Physics, Faculty of Fundamental Problems of Technology, Wroc{\l}aw University of Science and Technology, W. Wyspia{\'n}skiego 27, 50-370 Wroc{\l}aw, Poland\\
$^2$Institute of Physics, Polish Academy of Sciences, Al. Lotnik\'ow 32/46, 02-668 Warsaw, Poland\\
}
\begin{abstract}
We study theoretically the emission of dark solitons induced by a moving defect in a nonresonantly pumped exciton-polariton condensate. The number of created dark solitons per unit of time is found to be strongly dependent on the pump power. We relate the  observed dynamics of this process to the oscillations of the drag force experienced by the condensate. We investigate the stability of the polariton quantum fluid and present various types of dynamics depending on the condensate and moving obstacle parameters. Furthermore, we provide analytical expressions for dark soliton dynamics using the variational method adapted to the non-equilibrium polariton system. The determined dynamical equations are  found to be in excellent agreement with the results of numerical simulations.
\end{abstract}
\maketitle
\section{\label{sec:INTRO}INTRODUCTION}
Exciton-polaritons are bosonic quasi-particles formed as a superposition between a photon mode and a quantum well exciton, that exist in the strong coupling regime in semiconductor microcavities~\cite{Hopfield_1,Weisbuch_1,Kavokin_1}. The increased interest in exciton-polaritons is caused by their unusual properties, such as extremely low effective mass and strong exciton-mediated interparticle interactions~\cite{Carusotto_1,Deng_1}. The low effective mass allows for the creation of Bose-Einstein condensates at temperatures much higher than in the case of ultracold atomic gases~\cite{Christopoulos_1,Kasprzak_1}. In some material configurations, the critical condensation temperature can be higher than room temperature~\cite{Kasprzak_1,Christopoulos_1, Cohen_1,Guillet_1,Li_1}. Exciton-polaritons are also characterized by a finite lifetime. Due to the escape of photons through the mirrors of the microcavity, condensates need to be continuously pumped by an external laser pump or electrical contacts to compensate for the decay of particles. Moreover, polariton condensates can behave as a superfluid, which was observed in open-dissipative systems~\cite{Amo_1,Lerario_1}.

These exceptional properties of polariton quantum fluids bring posibility to observe topological and nonlinear excitations such as solitons and quantized vortices~\cite{Ostrovskaya_1,Lagoudakis_1,Sich_2,Amo_1}. Additionally, recent theoretical work predicted the existence of solitary waves in polariton topological insulators~\cite{Malpuech_1,Kartashov_2}. 
Diversity of nonlinear effects provides that polariton condensates are a promising area for the investigation  of quantum nonlinear photonics. In this work we focus on one of the types of nonlinear excitations which is a dark soliton. 

Dark soliton is a nonlinear excitation created when the effect of nonlinearity compensates the dispersion present in the system. It has the form of a localized density dip on a continuous wave background~\cite{Frantzeskakis_1}. The density dip in wave function separates regions with the same amplitude but different phases~\cite{Kevrekidis_1,Kivshar_1,Kivshar_2}.  This nonlinear wave is a fundamental excitation observed in many physical systems~\cite{Infeld_1}. It should be noted that experimentally there is a high degree of control over the polariton system. Moreover nonlinear excitations can be created optically or by defects already existing within the sample, as already demonstrated in multiple studies~\cite{Sich_1}. 

Nonlinear excitations in polariton quantum fluids have been studied in many theoretical and experimental works. Historically the first experimental observations of dark and bright  solitons were presented respectively in~\cite{Amo_1} and~\cite{Sich_2}. These works initiated extensive research in this field. Over the last few years research on polariton solitons led to observation of new phenomena such as generation of gap-solitons in one dimensional periodically modulated microwires~\cite{Tanese_1} and oblique half-dark solitons~\cite{Flayac_1,Hivet_1}. 
However, generation of dark solitons in a nonresonantly pumped polariton condensate is still an experimental challenge. In this work, we have developed a theoretical description of this problem, which may facilitate the experimental realization.

Many previously reported experimental results
 are accurately modeled by the Gross-Pitaevskii equation (GPE). The use of GPE makes it possible to describe fundamental properties of dark and bright solitons in a polariton condensate, such as their dynamics, stability, and continous emission~\cite{Smirnov_1,Pinsker_1,Pinsker_2,Pinsker_3,Xue_1,Chen_1,Kartashov_1,Piegon_1}. Successful creation of nonlinear excitations and their manipulation gave rise to a new concept of information processing based on vortices and soliton dynamics~\cite{Cancellieri_1, Ma_1,Goblot_1}. It is important to note that using dark solitons or quantized vortices for information processing devices requires a precise description of nonlinear dynamics, which is the aim of our work.

We describe the dynamics of dark solitons in a quasi-one dimensional nonresonantly pumped polariton condensate. In contrast to previous works~\cite{Larre_1,Kamchatnov_1,Kamchatnov_2,Kamchatnov_3,Larre_2,Tercas_1}, we take into account the effects of the hot uncondensed reservoir. We consider a dark soliton train generated by a defect moving with a constant velocity,  which can be created by an aditional off-resonant laser beam via the dynamic Stark effect~\cite{Hayat_1}. Both the soliton creation process and subsequent dynamics are analyzed in detail with various physical parameters. We determine the conditions necessary for soliton creation and link them to the analytical condition for stability of the polariton condensate. In addition, we determine the analytical form of soliton trajectories in the variational approximation, which turns out to reproduce numerical results very accurately. While the final formulas are identical to those obtained by the perturbation method~\cite{Smirnov_1}, the derivation within the variational approach is more transparent and easier to implement in the general case. Finally, we discuss the realation of dark solitons generated by the defect to exact Bekki-Nozaki hole solutions of the complex Ginzburg-Landau equation.

The paper is structured as follows. In Sec.~\ref{sec:MODEL}  we present a detailed description of the system under consideration. Next, we describe the model based on the Open-Dissipative Gross-Pitaevskii equation and present some relevant polariton condensate properties. In Sec.~\ref{sec:GENERATIONoDS} we  show examples of dark soliton generation and their continuous emission based on numerical simulations. We present drag force oscillations and show how the number of emitted solitons and oscillation period depend on the pump field intensity.  The influence of the condensate stability on the dynamics of generated dark solitons are investigated as well. In Sec.~\ref{sec:DARK_SOLITON_DYNAMIC} we present theory based on the adiabatic approximation and variational approximation for dark solitons~\cite{Kivshar_1,Kivshar_2} adapted to our model. We derive analytical equations for dark soliton trajectory, velocity and acceleration, and compare them to numerical simulations. In Sec.~\ref{sec:Bekki} we discuss the similarities and differences between tanh-shaped dark soliton solutions an Bekki-Nozaki holes. In  Sec.~\ref{sec:CONCLUSION} we summarize our work.
\section{\label{sec:MODEL}MODEL}

\begin{figure}
\includegraphics[width=0.5\textwidth]{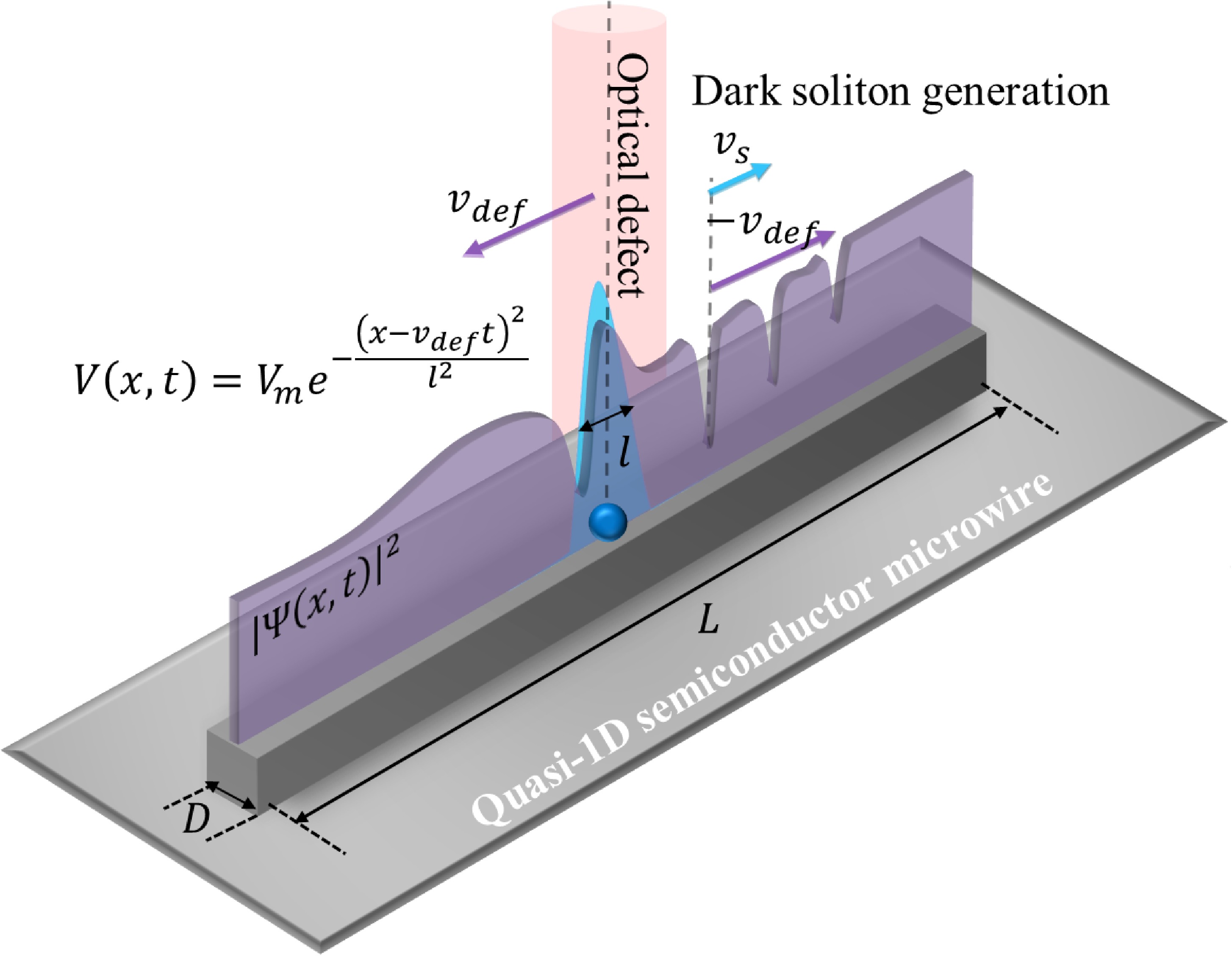}
\caption{Schematic presentation of the considered system. A quasi one-dimensional semiconductor microwire provides an effective spatial confinement. Polaritons are generated optically using a non-resonant excitation. The second laser beam, localized and detuned from the polariton energy, creates a dynamical optical defect moving with a constant velocity. The optical defect is due to the dynamic Stark shift.}
\label{fig:Fig_1}
\end{figure}

\subsection{\label{subsec:OBSTACLE_FLOW} Moving obstacle in a quasi-one-dimensional polariton condensate}
In the present work we consider a quasi-one dimensional semiconductor microcavity created by modern semiconductor technology (eg.~molecular beam epitaxy). This method allows the growth of structured photonic microstructures. Using the ion etching techniques one can create structures enabling photon and exciton confinement in a one-dimensional microwire, see Fig.~\ref{fig:Fig_1}. As demonstrated in~\cite{Hayat_1}, the dynamic Stark effect (DSE) with a far detuned laser beam allows for the creation of a dynamical potential, and in particular, an optically induced defect. The DSE creates an effective potential for the polaritons without the excitation of the exciton reservoir. In our work we modeled this effect introducing effective optical potential \(V(x,t)\) acting only on the polariton wave function, without additional excitation of the reservoir density. Perturbation of the polariton field caused by DSE is a possible method of dark or bright soliton generation~\cite{Zhang_1}. The change in the spatial position of the laser spot responsible for the creation of the defect could be performed with a spatial light modulator (SLM) or a rapidly tilting mirror, which driven by a piezoelectric element or an electrically controlled screw.

We model the  defect potential with a Gaussian function
\begin{equation}
V(x,t)=V_m e^{-\frac{(x-v_{def} t)^2}{l^2}},
\end{equation}
where \(V_m\) is the amplitude of the potential, \(l\) is an effective width of the Gaussian profile, and \(v_{def}\) is the obstacle velocity. Position of the optical defect is controlled by the detuned laser source. In the considered case, the created potential strains the polariton quantum fluid which allows for the observation of interesting behaviour and nonlinear effects such as Cherenkov radiation, superfluidity  or continuous soliton generation. Similar configuration has been used in recent works~\cite{Hakim_1,Nova_1,Pavloff_1,Larre_1,Kamchatnov_1,Kamchatnov_2,Abdullaev_1,Frisch_1,Engels_1,Bernard_1,Leszczyszyn_1} and black-hole analog laser configuration~\cite{Nova_1,Recati_1,Finazzi_1}. However, we consider here the configuration with a stationary condensate and a moving defect, in contrast to the fixed defect and a flowing condensate. This configuration is characterized by a larger region of stability~\cite{Natalia_1}. A conservative quantum fluid, such as atomic Bose-Einstein condensate, described by the Nonlinear Schrodinger Equation, is characterized by two different behaviours separated by the Landau criterion.  As shown in~\cite{Hakim_1}, effects observed in the system strongly depend on the critical condition for the flow velocity. Above the critical velocity, steady flow solution vanishes by merging with unstable solution in the usual saddle-node bifurcation. This unstable regime is marked by a continuous emission of gray solitons~\cite{Hakim_1,Abdullaev_1,Pham_1}.  As will be shown below, this behaviour can be also observed in a nonresonantly pumped polariton condensate, but only in the weak pumping regime, when \(P \approx P_{th}\), where \(P_{th}\) is the condensation threshold.

\subsection{\label{subsec:ODGPE}Open-dissipative Gross-Pitaevskii equation}
 We describe the considered system using the open-dissipative Gross-Pitaevskii equation (ODGPE)~\cite{Wouters_1}. This equation governs the time evolution of the complex polariton order parameter \(\Psi(x,t)\) and is coupled to the rate equation describing the uncondensed exciton reservoir \(n_R(x,t)\)
\begin{equation}\label{1}
i \hbar \frac{\partial\Psi} {\partial t} = -\frac{{\hbar}^2}{2m^{*}}\frac{\partial ^2 \Psi}{\partial x^2}+\frac{i}{2}(R^{1D} n_R -\gamma_C)\Psi+U\Psi+V\Psi,
\end{equation}
\begin{equation}\label{2}
\frac{\partial n_R} {\partial t} = P(x,t)-(\gamma _R+R^{1D}|\Psi|^2)n_R,
\end{equation}
where: \(m^*\)is the effective mass, \(P(x,t)\)  is the  exciton generation rate given by pumping laser profile, and \(R^{1D}\) is the stimulated scattering rate. We assume that the stimulated relaxation of polaritons from reservoir to the condensate is given by a  linear term \(R^{1D} n_R(x,t)\). In the above, \(U(x,t)\) is the effective potential determined by the blueshift caused by interactions
\begin{equation}\label{2}
U(x,t)=g_C|\Psi|^2+g_R n_{R},
\end{equation}
where \(g_{C}\) and \(g_{R}\) are the interaction coefficients describing the interactions between the condensed polaritons and between the reservoir particles and the condensate, respectively. The finite lifetime of polaritons and the reservoir \(\tau_C\), \(\tau_R\) are described by loss rates \(\gamma_C=\frac{1}{\tau_C}\) and \(\gamma_R=\frac{1}{\tau_R}\). It should be noted that when polaritons are confined in a one-dimensional semiconductor quantum wire, the system parameters must be rescaled as compared to the two-dimensional case. The nonlinear interactions coefficients and the stimulated scattering rate are \((R^{1D},g_{C,R}^{1D})=(R^{2D},g_{C,R}^{2D})/\sqrt{2 \pi D^2}\), where \(D\) stands for the width of the microwire, see Fig.\ref{fig:Fig_1}. 

\begin{figure}
\includegraphics[width=0.5\textwidth]{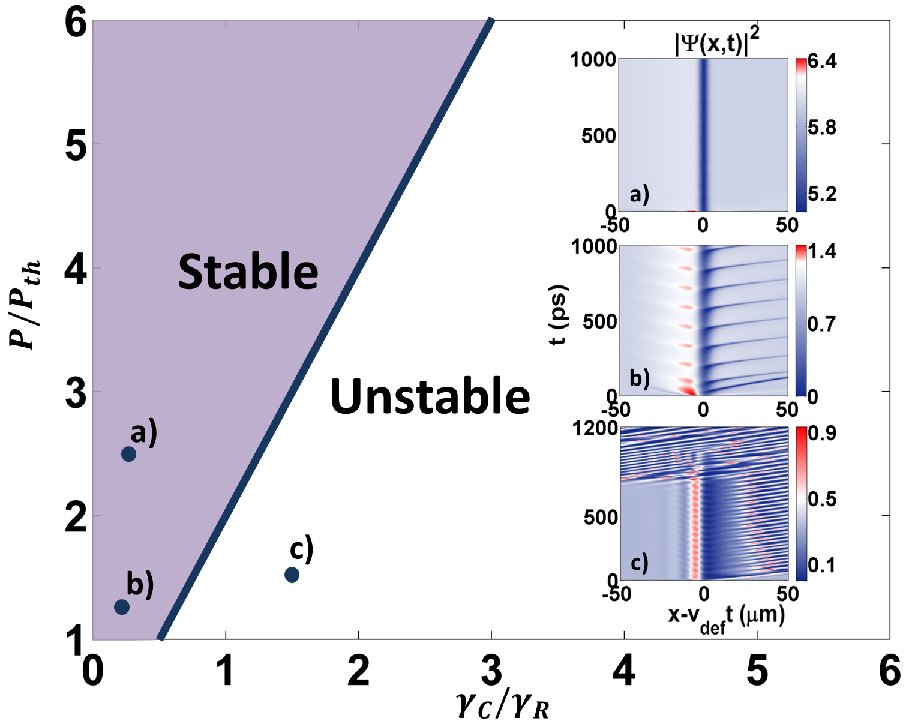}
\caption{Diagram of stability of a homogeneous condensate and the corresponding dynamics with a moving defect. Solid line represents the stability criterion (8) for \(g_R=2g_C\). Three insets show different types of evolution in stable and unstable cases. Note that the horizontal axis in the insets  corresponds to the position in a reference frame moving together with the defect. (a) Stable with no solitons, (b) weakly unstable with soliton emission, and  (c) strongly unstable with turbulent evolution. Parameters are  a) \(\frac{\gamma_C}{\gamma_R}=0.25,\) \(\frac{P}{P_{th}}=1.25\),
 b) \( \frac{\gamma_C}{\gamma_R}=0.8\), \(\frac{P}{P_{th}}=1.5\), 
 c) \( \frac{\gamma_C}{\gamma_R}=2\), \(\frac{P}{P_{th}}=1.5\),
 and  \(R=1,\ V_m=0.8,\ v_{def}=0.6,\ g_C=1,g_R=2 g_C,\ l=1\).
}
\label{fig:Fig_2}
\end{figure}

It is useful to introduce dimensionless parameters in the ODGPE system~\cite{Natalia_1}. We introduce dimensionless space, time, wave function and material coefficients according to \(x=\xi \widetilde{x},\ t=\tau \widetilde{t},\ \Psi=(\xi \beta)^{-1/2}  \widetilde{\Psi}, \ n_R=(\xi \beta)^{-1} \widetilde{n}_R, \ R^{1D}=(\xi \beta / \tau)\widetilde{R}^{1D}, \ (g_C^{1D},g_R^{1D})=(\hbar \xi \beta / \tau)(\widetilde{g}_C \widetilde{g}_R), \ (\gamma_C,\gamma_R)=\tau^{-1}(\widetilde{\gamma}_C,\widetilde{\gamma}_R),\ P(x)=(1/\xi \beta \tau)\widetilde{P}(x)\), where \(\xi=\sqrt{\frac{\hbar \tau}{2 m^{*}}}\) and \(\tau\), \(\beta\) are arbitrary scaling parameters. 
Equations~(\ref{1}) and~(\ref{2}) take the form
\begin{equation}\label{4}
i \frac{\partial\Psi} {\partial t} =
\Big[-\frac{\partial ^2}{\partial x^2}+\frac{i}{2}(R n_R -\gamma_C)+g_C|\Psi|^2+g_R n_R+V(x,t)\Big]\Psi,
\end{equation}
\begin{equation}\label{5}
\frac{\partial n_R} {\partial t} = P(x,t)-(\gamma_R+R|\psi|^2)n_R.
\end{equation}
In the case of homogeneous continuous wave pumping, \(P(x,t)=P=\textrm{const}\), the uniform condensate solution is
\begin{align}
\Psi(x,t)=\Psi _0 e^{-i\mu _0 t}, \\
\nonumber n_R(x,t)=n^0_R.
\end{align}
When loss and gain are balanced, the equilibrium densities and the chemical potential of the condensate are given by
\( \mu_0 =g|\Psi _0|^2+g_R n^0_R\) and \( n^0_R=n^{th}_R=\frac{\gamma _C}{R}
\). Above the condensation threshold,  when \(P>P_{th}\), the condensate density takes the form
\begin{equation}
|\Psi _0|^2 =\frac{P-P_{th}}{\gamma_C}.
\end{equation}
The stability of a homogeneous condensate depends on system parameters. The analytical condition for stability was  found to be \cite{Smirnov_1}
\begin{equation} \label{stab_lim}
\frac{P}{P_{th}}>\frac{g_R}{g_C}\frac{\gamma _C}{\gamma _R}.
\end{equation}
When this condition is not fulfilled, the condensate becomes dynamically (modulationally) unstable. In this case, small fluctuations grow exponentially in time, and dramatically affect the time evolution of the condensate. The stability diagram is presented in  Fig.~\ref{fig:Fig_2}. The solid line, corresponding to the formula above, separates the stable and unstable regions in parameter space.

\section{\label{sec:GENERATIONoDS}GENERATION OF DARK SOLITONS}

\subsection{\label{subsec:DARK_SOLITON} Analytical form of a dark soliton}

It is instructive to recall properties of dark solitons in the case of the Nonlinear Schr\"odinger equation (NLSE), which in the context of Bose-Einstein condensation is also named the Gross-Pitaevskii equation (GPE)
\begin{equation} \label{NLSE}
i\frac{\partial \psi}{\partial t}+\frac{1}{2}\nabla^2_x\psi +\Gamma (1-|\psi|^2)\psi=0,
\end{equation}
where \(\Gamma\) is the nonlinearity parameter.
The GPE admits nontrivial dark soliton solutions in the form~\cite{Kivshar_1,Kivshar_2,Kevrekidis_1,Theocharis_1,Frantzeskakis_1}
\begin{equation}
\psi_s=i A+B \tanh(D \xi(x,t)),
\end{equation}
where \(\xi(x,t)=x-x_0(t)\) takes into account the motion of the soliton center \(x_0(t)=v_st\) and \(v_s\) is the relative velocity between the soliton and the stationary background. Parameters \(A\) and \(B\) are related by a simple trigonometric relation \(A^2+B^2=1\). We can write \(A=\sin(\phi)\) and \(B=\cos(\phi)\) where \(\phi\) is the soliton phase angle \(\big( \phi<\frac{\pi}{2}\big)\). The phase jump across the dark soliton \(\Delta \phi\) is given by the relation  \(\Delta \phi=2\big[\tan^{-1}(\frac{B}{A})-\frac{\pi}{2}\big]=-2\tan^{-1}(A/B)\). In the case of unperturbed GPE, one has \(\frac{d x_0(t)}{dt}=v_s=A\) and \(D=B\). In the perturbed case considered below, these relations may not be valid~\cite{Kivshar_1,Kivshar_2,Frantzeskakis_1}. In the limiting case, when\(B=1\), Eq.~{(20)} describes a static black soliton with velocity \(v_s=0\). In this case, the soliton phase shift is exactly zero, and the phase is given by the Heaviside function of an amplitude equal to \(\pi\). Otherwise, when \(B \ll 1\), soliton is moving with a nonzero velocity dependent on the phase shift. This type of soliton is called a grey soliton. Dark soliton effectively behaves like a classical particle, obeying equation of motion and can be described within the classical mechanics theory~\cite{Kivshar_1,Kevrekidis_1,Frantzeskakis_1}.

Gross-Pitaevskii equation is  Galilean invariant, which means that the motion of a soliton is the same in an inertial frame, when condensate is set into motion with background velocity $v_b$. The soliton velocity can be then expressed \(v_s'=v_s+v_b\), where \(v_s'\) is the soliton velocity with respect to the moving condensate. 
When such condensate flow is considered, equation (20) takes the form
\begin{equation}
\psi_s=\psi_0 e^{i v_b x-i\mu t}\Big[i A+B\tanh\big(D\xi(x,t)\big)\Big],
\end{equation}
Taking into account the relation between the soliton velocity and amplitude, we can write the full solution 
\begin{equation} \label{eq:solitonFlow}
\psi_s=\psi_0 e^{i v_b x-i\mu t}\Big[i v_s+\sqrt{1-v_s^2}\tanh\big(\sqrt{1-v_s^2}(x-x_0t)\big)\Big],
\end{equation}

Note that in an exciton-polariton condensate, dark solitons are subject to dissipation. Dissipation leads to the gradual decrease of the soliton amplitude related to the B parameter, which consequently leads to soliton acceleration~\cite{Smirnov_1}. This behaviour will be evident in the solutions of the full problem~(\ref{1})-(\ref{2}) considered below.

\begin{figure}
\includegraphics[width=0.5\textwidth]{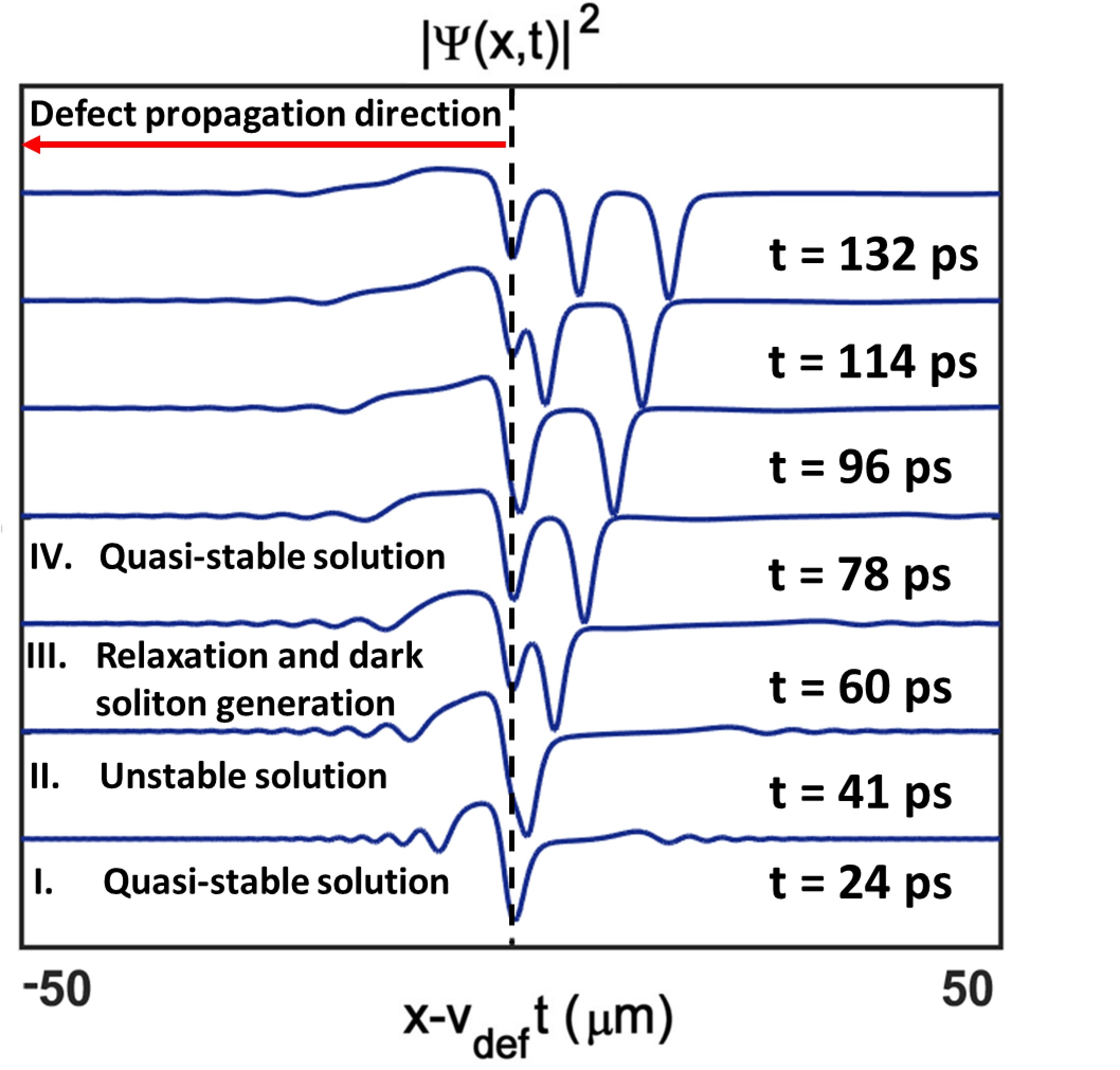}
\caption{The process of dark soliton generation. At \(t=24 ps\) the condensate is in a quasi-stable regime, with a density dip at the center of the defect (dashed black line). Next, due to the movement of the defect, the density dip is released from the pinning potential and becomes a dark soliton. At \(t=60 ps\) dark soliton is emitted away completely and the unstable solution decays into the quasi-stable again. The sequence is repeated multiple times.}
\label{fig:Fig_7}
\end{figure}

\subsection{\label{subsec:NON_EQULIBRIUM_FLOW} Generation of solitons by a moving obstacle}
For the modeling of the system we used a numerical scheme based on the fourth-order Runge-Kutta spectral method with periodic boundary conditions.
We present numerical solutions of the full system of time evolution equations~(\ref{1})-(\ref{2}) in Fig.~\ref{fig:Fig_2}, which shows the diagram of stability of the homogeneous condensate. The three insets show the typical dynamics in the presence of a moving defect. Note that the horizontal axis corresponds to the position in a reference frame moving together with the defect, and its position is fixed at $x-v_{def} t=0$. Three different regimes can be clearly distinguished depending on the pumping power and the decay parameters. At high power and in the regime of a stable background, the density is practically undisturbed except for the local density dip due to the defect potential, as in Fig.~\ref{fig:Fig_2}(a). The condensate behaves as a superfluid. At lower pump powers, closer to the threshold, a nearly-periodic train of dark solitons is created, which propagate away from the defect, see Fig.~\ref{fig:Fig_2}(b). This is accompanied by the oscillations of the condensate density in the vicinity of the defect. Finally, in the regime of unstable condensate background, shown in Fig.~\ref{fig:Fig_2}(c), the oscillatory dynamics are destroyed by the dynamical instability at later times of the time evolution.

We would like to emphasize that due to the lack of Galilean invariance of the system~(\ref{1})-(\ref{2}), the setup with a moving obstacle and stationary condensate considered here is not equivalent to the case of condensate flowing past a stationary defect~\cite{Amo_2}. In contrast to the pure GPE case, Eq.~\ref{NLSE}, the solutions of a condensate moving with velocity $v_b$ do not have corresponding solutions in a stationary condensate, due to the presence of a immobile reservoir. Indeed, it was demonstrated~\cite{Natalia_1} that the stability region in parameter space shrinks considerably in the presence of condensate flow.

The results of Fig.~\ref{fig:Fig_2} suggest that the optimal region in parameter space for the investigation of soliton dynamics is the  stable region at low pumping power (b). This is exactly the region which becomes unstable in a moving condensate with a stationary reservoir.  We refer the reader to~\cite{Natalia_1} for more details on the stability of the flowing condensate. Thus the configuration with a moving optical defect and a stationary condensate is much more favorable for the generation of dark solitons in a nonresonantly pumped condensate than the one with condensate flowing past a defect.

\begin{figure}
\includegraphics[width=0.5\textwidth]{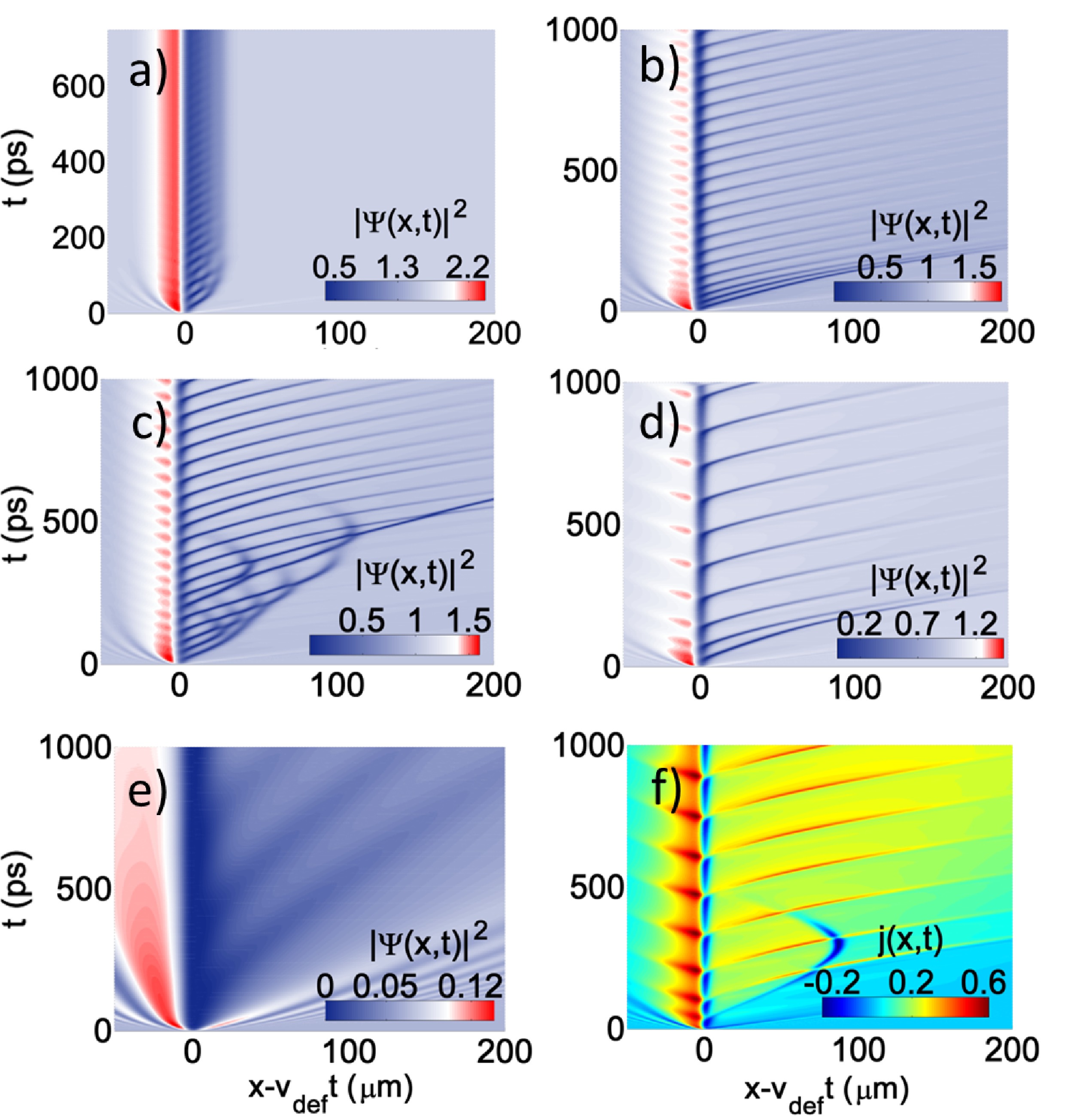}
\caption{(a)-(e) Examples of typical evolution of polariton density. In panel (c) the parameters in dimensionless units are \(\gamma_R=6.66\), \(\gamma_C=0.15 \gamma_R\), \(P=1.15P_{th}\)  and  \(R=1,\ V_m=0.8,\ v_{def}=0.6\ \  (=0.42{c_s}),\ \ g_C=1,g_R=2 g_C,\ l=1\). The other graphs differ with respect to (c) in a single parameter, (a) $P=1.25 P_{th}$,
(b) $V_m=1.3$,
(d) $v_{def}= 0.43 c_s$,
(e) $P=1.01 P_{th}$,
(f) $l= 0.7$.
(f) Map of the polariton density current.
}
\label{fig:Fig_3}
\end{figure}

The process of soliton creation by the potential defect has been analyzed and described in several works ~\cite{Hakim_1,Abdullaev_1,Pham_1,Bernard_1,Leszczyszyn_1}. As was predicted in the case of one dimensional defocusing nonlinear Schrodinger equation (conserative case) in the case of a flow past a potential obstacle two types of solutions exist, a stable and an unstable one~\cite{Hakim_1,Abdullaev_1,Pham_1}. The two types of condensate flow coalesce through a saddle-node bifurcation, which can be found analytically using the hydrodynamic approximation~\cite{Hakim_1}.  The stable steady state solution was obtained in the system up to the characteristic critical velocity, which depends on the potential \(V(x,t)\)~\cite{Hakim_1}.  In this case, the stable solution was localized at the obstacle position. 
Above the critical velocity, the two types of solutions merge. In this case, no steady solution exists in the system and dark solitons are repeatedly emitted. It should be noted that after emission of a soliton the system tries to relax to the stable solution emitting gray solitons and returning to the quasi-steady solution. This unstable solution is an example of transitional state. 
The quasi-steady solutions are perturbed by the current density flowing across the obstacle. When the barrier is exceeded, a soliton is emitted. Next, the system is in the quasi-stationary state and process repeats again. 
Figure~\ref{fig:Fig_7} present schematically the process of soliton generation which occurs along these lines. While our system is driven dissipative and does not allow for the analytical solutions, the main phases of the soliton generation process are the same as in the conservative case.

In Fig.~\ref{fig:Fig_3} we analyze the dynamics of soliton creation in the cases when some of the parameters of the system are varied. The case shown in panel (c) is our reference solution, which shows the generation of a soliton train. The transient dynamics at early times are due to the gradual buildup of condensate phase and density pattern after switching on the moving condensate potential at $t=0$. Several first solitons are generated with a velocity such that their trajectories bend back towards the defect (see Sec.~\ref{subsec:LAGRANGIAN_APPROACH} for more details on the soliton trajectory). The backward bending is due to the gradual appearance of the defect-induced global density current, which gives rise to  the background condensate velocity as in Eq.~(\ref{eq:solitonFlow}). The backward bending indicates that the initial soliton velocity is in fact negative (points towards the defect) in the reference frame moving together with the condensate background, see Sec.~\ref{subsec:LAGRANGIAN_APPROACH}. At later times, solitons are created periodically with an initial relative velocity that is positive, while their trajectories are pointing outwards. We find that this transient behaviour is strongly dependent on system parameters, in particular on the depth of the defect potential. In panel (b) the defect depth was increased, which led to the more regular evolution with a shorter transient time. This can be understood as an effect of a increased stiffness of the wavefunction near the deep defect, thanks to which the condensate regular dynamics is established more quickly.

In panels (a) and (e) we demonstrate the effect of the variation of pumping power. In panel (a), stronger pumping than in panel (c) leads to the decrease of the distance in time between the subsequent solitons, but also shortens the soliton lifetime. Note that the lifetime is limited due to dissipative nature of polaritons~\cite{Smirnov_1}. Since by increasing the pumping power one is moving towards the "more stable" region in the phase diagram (see Fig.~\ref{fig:Fig_2}), the shortening of soliton lifetime can be understood as the effect of increased stability of a homogeneous, soliton-free solution. This observation is confirmed by the analytical prediction for soliton trajectory, Eq.~(\ref{33}). The increase of soliton emission rate is, on the other hand, related to the increase of polariton density and consequently, the shortening of the nonlinear timescale.
In the low density case (e) no solitons are created due to the lack of sufficiently strong nonlinearity. Only some linear waves are seen in this panel, which undergo diffraction without well defined density dips characteristic for solitons. The absence of solitons in this case is natural as the  nonlinearity is insufficiently strong. Finally, in panel (d), we show the dynamics in the case of a slower defect velocity, which  results in a longer time interval between soliton creation events. The reduced emission rate is due to clearly visible slower buildup of the polariton density dip, which is the precursor of soliton emission.

\subsection{\label{subsec:SUPERFLUIDITY_BRAKE_DOWN} Breakdown of superfluidity and drag force oscillations}

\begin{figure}
\includegraphics[width=0.5\textwidth]{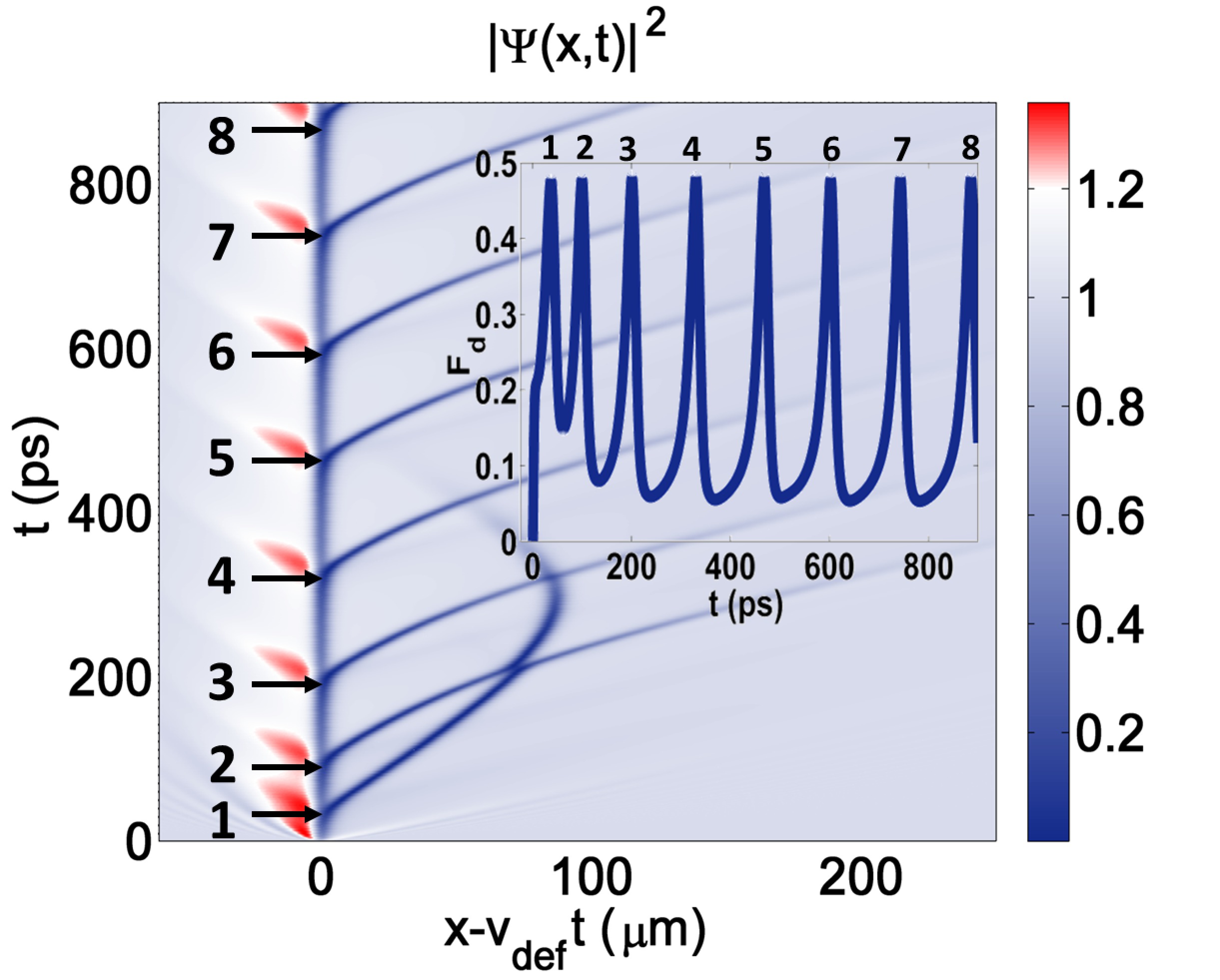}
\caption{Emission of dark solitons and the corresponding drag force oscillations (inset, \(F_d\) is given in dimensionless units). Maximum drag occurs at the time when solitons are generated by the defect. 
Parameters are \(R = 1, \frac{\gamma_C}{\gamma_R} = 0.15, \frac{P}{P_{th}} = 1.15, V_m = 0.8, v_{def} = 0.6 = 0.43 c_s, g_C = 1, g_R = 2g_C , l = 0.7\). Corresponding parameters in physical units are \(\tau_C = 3 ps\), \(\tau_R=0.45 ps\), \(g_R^{1D}=17.39 \times 10^{-3} \mu eV \mu m^2\), \(g_C^{1D}=8.69 \times 10^{-3}  \mu eV \mu m^2\), \(R^{1D}=13.21 \times 10^{-3} \mu m^2 ps^{-1}\), \(\gamma_R=2.22 ps^{-1}\), \(\gamma_C=0.33 ps^{-1}\), \(P_{th}=281 \mu m ^{-2} ps^{-1}\). }
\label{fig:Fig_4}
\end{figure}

In order to investigate quantitatively the breakdown of superfluidity and emission of dark solitons, we determine the drag force \(F_d\) exerted on the moving obstacle \cite{Larre_1,Regemortel_1}:
\begin{equation}
F_d=\int_{-\infty}^{\infty}|\psi(x,t)|^2\frac{\partial}{\partial x}V(x,t)d x .
\end{equation}
Landau criterion for superfluidity is broken when the local velocity of particle current exceeds the value of the critical velocity. The flow in vicinity of the obstacle becomes dissipative, leading to the increase of the drag force and the appearance of excitations induced by the defect. 

We observe analogous behaviour in the nonequilibrium system of exciton-polariton condensate. Drag force oscillations and the corresponding soliton emission are presented in  {Fig.~\ref{fig:Fig_4}}. Drag force increases gradually in each period of oscillations and is sharply peaked at the instants that correspond exactly to the appearance of a new soliton. The calculated time evolution of drag force can therefore be used to determine the time instants when dark solitons are created for various condensate parameters. 

\begin{figure}
\includegraphics[width=0.5\textwidth]{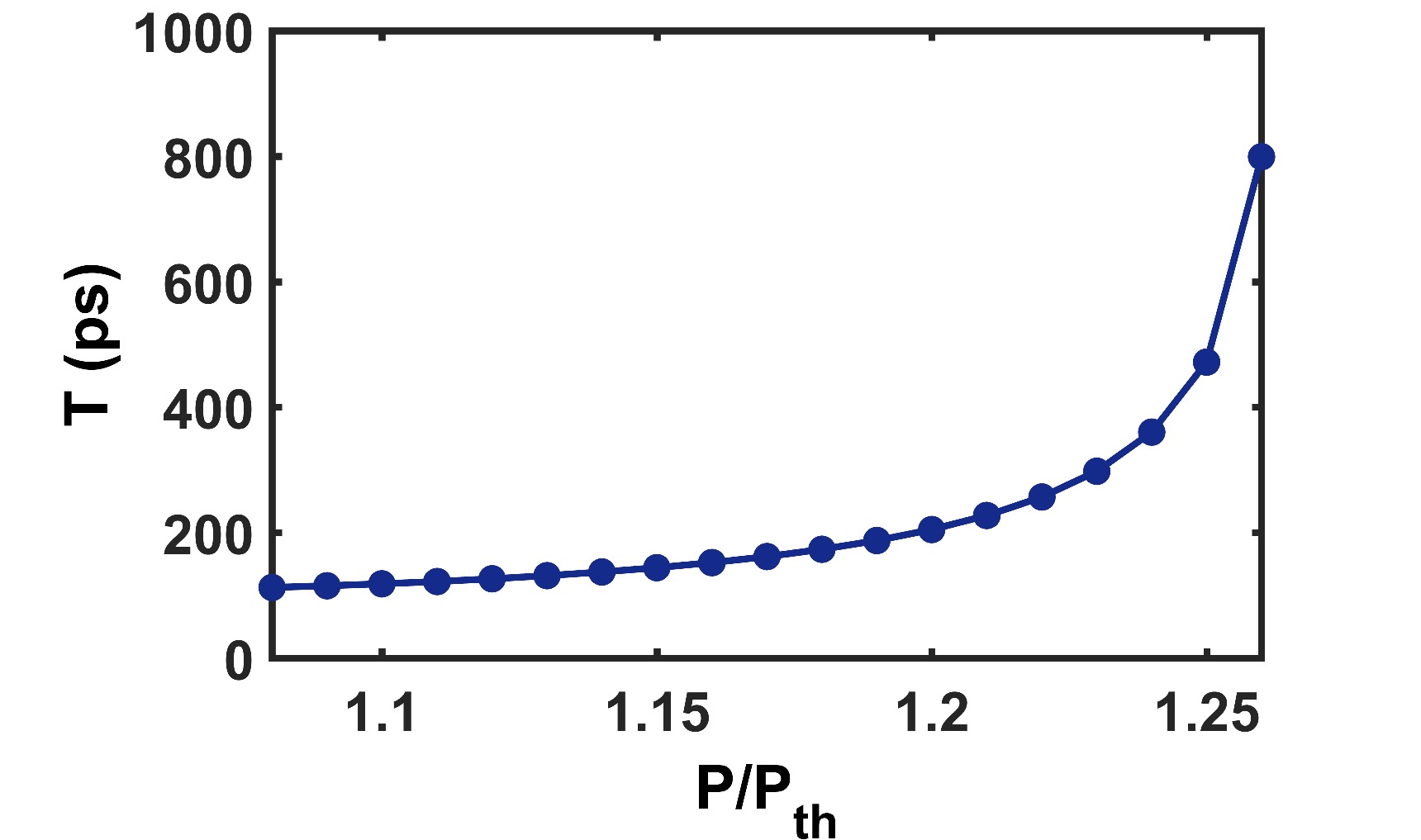}
\caption{The period between generation of subsequent solitons $T$ after its stabilization, as a function of pump power.  
Simulation parameters are the same as in  Fig.~\ref{fig:Fig_4}.}
\label{fig:Fig_5}
\end{figure}

Based on analysis of the drag force dynamics, we can extract the characteristics of soliton train formation. Fig.~\ref{fig:Fig_5} presents dark soliton emission time as a function of pump intensity. The time interval between subsequent solitons tends to a constant value, different at each power, which shows that the system dynamics becomes eventually periodic after an initial transient time. This is related to stabilization of density currents created in the condensate: \(j(x,t)=\frac{i}{2}\big( \Psi^* \frac{\partial \Psi}{\partial x}  -\frac{\partial \Psi^*}{\partial x}\Psi \big)\) presented on panel (f) in Fig.~\ref{fig:Fig_3}.
Moreover, close to the condensation threshold dark soliton emission period is shorter than in the case of more intensive excitation eg. at~\(P=1.26 P_{th}\).  
At very high pumping soliton emission disappears completely and the condensate becomes stable.
Note that in the presented simulations the defect velocity \(v_{def}=0.6\) is the same at each power, while sound velocity increases with pump intensity. In result, the defect velocity relative to sound velocity changes from \(v_{def}=0.32c_s\)  to \(v_{def}=0.73 c_s\) where $c_s$ is the sound velocity away from the defect. However, the local sound velocity in the vicinity of the defect is strongly reduced due to the dip in the condensate density, which leads to the breakdown of superfluidity even at \(v_{def}< c_s\).

\section{\label{sec:DARK_SOLITON_DYNAMIC} Analytical prediction of soliton trajectories}

\subsection{\label{subsec:ADIABATIC_APPROX}Adiabatic approximation in the weak pumping limit}

In the following, we will consider the adiabatic approximation, which assumes that the reservoir density quickly adjusts to its steady-state value, \(\frac{\partial n_R}{\partial t}=0\). This assumption is justified under certain conditions~\cite{Natalia_2}, in particular when the ratio of condensate to reservoir lifetimes is large, i.e.~$\gamma_R/\gamma_C \gg 1$. The reservoir density adiabatically follows the change of macroscopic polariton wave function. In this case the reservoir density  takes the form
\begin{equation}\label{14}
n_R(x,t)=\frac{P}{\gamma_R+R|\Psi(x,t)|^2},
\end{equation}
  The limit of the validity of adiabatic approximation is estimated by three conditions~\cite{Natalia_2} \(k^2 \ll 2m^*/(\hbar \tau_R)\), \(g_C<R\) and \(  P/P_{th} \gg \frac{g_R}{R}\frac{\gamma_C}{\gamma_R}\). Under this assumption reservoir density \(n_R\) is able to quickly adjust to the condensate density \(|\Psi(x,t)|^2\). 

 To determine temporal evolution of dark solitons, we  consider the limit of pumping power slightly above threshold, $P\gtrsim P_{th}$, in which generation of solitons is observed in simulations, see Fig.~\ref{fig:Fig_3}. Following~\cite{Smirnov_1}, we introduce small perturbations in the form of the slowly varying condensate wave function envelope \(\psi(x,t)\), close to unity which corresponds to a homogeneous steady state, and the reservoir density perturbation \(m_R(x,t)\)
\begin{equation}\label{15}
\Psi(x,t)=\Psi_0\psi(x,t)e^{-i\mu_0 t},
\end{equation}
\begin{equation}\label{16}
n_R=n_R^0+m_R(x,t).
\end{equation}
We expand formula~(\ref{14}) in Taylor series up to second order in \(|\Psi(x,t)|^2\) and use the formula~(\ref{16})
\begin{equation}\label{17}
n_R^0+m_R(x,t) = \frac{P}{\gamma_R}+\frac{P R}{\gamma_R^2}|\Psi(x,t)|^2,
\end{equation}
After some algebra, the above equation allows to obtain an analytical formula for the reservoir perturbation \(m_R(x,t)\)
\begin{equation}\label{18}
m_R(x,t)=\frac{\gamma_C}{\gamma_R}\Psi_0^2\Big( 1-\frac{P}{P_{th}}|\psi(x,t)|^2)\Big).
\end{equation}
Substituting equation (\ref{15}) and (\ref{18})  into (\ref{5}) we obtained the dynamical equation for wave function perturbation
\begin{equation} \label{19}
\begin{split}
i \frac{\partial\psi} {\partial t} & =-\frac{1}{2}\frac{\partial ^2 \psi}{\partial x^2} + \frac{i}{2}R m_R \psi \\
 & + \Big(g_C\Psi_0+g_R n_R^0\Big(1-\frac{P}{P_{th}}\Big)\Big)\Big(1-|\psi|^2\Big)\psi,
\end{split}
\end{equation}
where the second term represents the effective nonlinearity in the system. This equation describes time evolution of a small perturbation of the condensate. 
As we consider the regime of a weak pumping, $P\gtrsim P_{th}$, Eqs.~(\ref{18}) and (\ref{19}) are simplified to
\begin{equation}\label{21}
\begin{split}
m_R(x,t) =
 \frac{\gamma_C}{\gamma_R}\Psi_0^2\Big( 1-|\psi(x,t)|^2)\Big),
\end{split}
\end{equation}
\begin{equation}\label{22}
  i \frac{\partial\psi} {\partial t}+
\frac{1}{2}\frac{\partial ^2 \psi}{\partial x^2} - g_c\Psi_0^2(|\psi|^2-1)\psi -\frac{i}{2}R m_R \psi =0.
\end{equation}
The above set of equations is equivalent to the well known complex Ginxburg-Landau eqaution (CGLE)~\cite{Popp_FromDarkSolitons}.

\subsection{\label{subsec:LAGRANGIAN_APPROACH}Variational approximation}

In this subsection we will analyze the dynamics of small perturbations, Eqs.~(\ref{21}),~(\ref{22}) within the Lagrangian formalism to describe dark soliton evolution. In contrast to the systematic perturbation method considered in~\cite{Smirnov_1}, the variational approximation used here provides a transparent and straightforward, easy to generalize method for derivation of nonlinear wave dynamics, at the expense of somewhat less precisely defined assumptions. It is instructive to compare the two methods, which, as we will show below, give identical analytical formulas for dark soliton trajectories. 

For clarity, we begin our presentation with recalling the approach presented in~\cite{Kivshar_1} in the context of the pure nonlinear Schr\"odinger equation, Eq.~(\ref{NLSE}).
This equation can be obtained from the Lagrangian density
\begin{eqnarray}\label{24}
\mathcal{L}&=&\frac{i}{2}\Big(\psi^*\frac{\partial \psi}{\partial t}-\psi\frac{\partial \psi^*}{\partial t}\psi\Big)\Big(1-\frac{1}{|\psi|^2}\Big)+\\\nonumber
&&-\frac{1}{2}\Big|\frac{\partial \psi}{\partial x} \Big|^2- \frac{1}{2} \Gamma \Big(|\psi|^2-1\Big)^2,
\end{eqnarray}

and the Euler-Lagrange equations
\begin{equation}\label{25}
\frac{\partial \mathcal{L}}{\partial u}-\sum_v \frac{\partial}{\partial v}\Big(\frac{\partial \mathcal{L}}{\partial u_v}\Big)=0,
\end{equation}
where \(v=\{x,t\}\) and \(u=\{\psi,\psi^*\}\). 
The Lagrangian of the system is the integral  of the Lagrangian density over space, \(L=\int_{-\infty}^{\infty}\mathcal{L} d x\). Assuming the dark soliton Ansatz in the form \(\psi=\psi_s=B\tanh\big(D(x-x_0)\big)+i A\) we obtain the ''variational'' Lagrangian
\begin{equation}\label{26}
L=2\frac{x_0}{dt}\Big[-AB +\tan^{-1}\Big(\frac{B}{A}\Big)\Big]-\frac{2}{3}\Big(B^2D+\Gamma\frac{B^4}{D}\Big).
\end{equation}
We now write down the Euler-Lagrange equations as
\begin{equation}\label{27}
\frac{\partial L}{\partial \alpha_j}-\frac{d}{d t}\Big(\frac{\partial L}{\partial \alpha^{'}_j}\Big)=0,
\end{equation}
where \(\alpha_j=x_0,A,B,D\) are the soliton parameters (which in general are functions of time) and \(\alpha_j{'}=\frac{d}{dt}\alpha_j\). Substituting (\ref{26}) to the Euler-Lagrange equation~(\ref{27}) we can derive evolution equations for the soliton parameters~\cite{Kivshar_1,Kivshar_2,Frantzeskakis_1,Kevrekidis_1,Pinsker_3}.

We note that similar approach based on Lagrangian formalism was successfully used to determine the parameters of stationary soliton solutions in spinor polariton condensates~\cite{Pinsker_3}.  In our work we used this formalism and perturbative theory to describe dark soliton dynamics.

Let us now return to the Eq.~(\ref{22}). The basic difference between Eq.~(\ref{22}) and (\ref{NLSE}) is the dissipative imaginary term \(\frac{i}{2}R m_R(x,t) \psi(x,t)\) related to the small reservoir perturbation. We can think of Eq.~(\ref{21}) as a generalized nonlinear Schrodinger equation with an added term \(\mathcal{R}(m_R,\psi)\) which is a functional of \(m_R(x,t)\) and \(\psi(x,t)\)
\begin{equation}\label{28}
  i \frac{\partial\psi} {\partial t}+
-\frac{1}{2}\frac{\partial ^2 \psi}{\partial x^2} - \Gamma(|\psi|^2-1)\psi = \mathcal{R}(m_R,\psi),
\end{equation}\label{29}
where \(\mathcal{R}(m_R,\psi)=\frac{i}{2} R  \frac{\gamma_C}{\gamma_R}\Psi_0^2\Big( 1-|\psi(x,t)|^2)\Big)\), and \(\Gamma=g_c\Psi_0\) in the weak pumping regime.
In this case evolution of soliton parameters can be obtained from the generalized Euler-Lagrange equations~\cite{Kivshar_1,Kivshar_2}
\begin{equation}\label{28}
\frac{\partial L}{\partial \alpha_j}-\frac{d}{d t}\Big(\frac{\partial L}{\partial \alpha^{'}_j}\Big)=2\Re\Big( \int_{-\infty}^{\infty} \mathcal{R}^*(\psi_s)\frac{\partial \psi_s}{\partial \alpha_j } d x\Big),
\end{equation}
where \(\Re\) is the real part of the expression. For instance, if we chose the parameter as \(\alpha_j=x_0\) we obtain the following result
\begin{equation}\label{30}
A(t)^{'}=\frac{4}{3} \frac{\gamma_C}{\gamma_R}R|\Psi_0|^2A(t)B(t).
\end{equation}
The above equation describes dynamics of the soliton phase. Using the above equation and the relations between soliton parameters we obtain equations for soliton acceleration, velocity and trajectory
\begin{equation}\label{31}
a_s(t)=\frac{d v_s}{d t}=\frac{1}{2 \tau_s}v_s(1-v_s^2),
\end{equation}
\begin{equation}\label{32}
v_s(t)=\int a_s(t) d t=\pm\sqrt{\frac{v_{s0}^2e^{\frac{t}{\tau_s}}}{1-v_{s0}^2+v_{s0}^2e^{\frac{t}{\tau_s}}}},
\end{equation}
\begin{equation}\label{33}
\begin{split}
& s_s(t)=\int v_s(t) d t=\\
& 2 \tau_s \tanh^{-1}\Big(\frac{1}{v_{s0}}\sqrt{-e^{-\frac{t}{\tau_s}}v_{s0}^2+v_{s0}^2+e^{-\frac{t}{\tau_s}}}\Big),
\end{split}
\end{equation}
where \(\tau_s=\frac{3}{2}\frac{\gamma_C}{\gamma_R R |\Psi_0|^2}\) is the dark soliton lifetime~\cite{Smirnov_1}. 
These  formulas are identical to these obtained from the perturbation theory presented in~\cite{Smirnov_1}.

\begin{figure}
\includegraphics[width=0.5\textwidth]{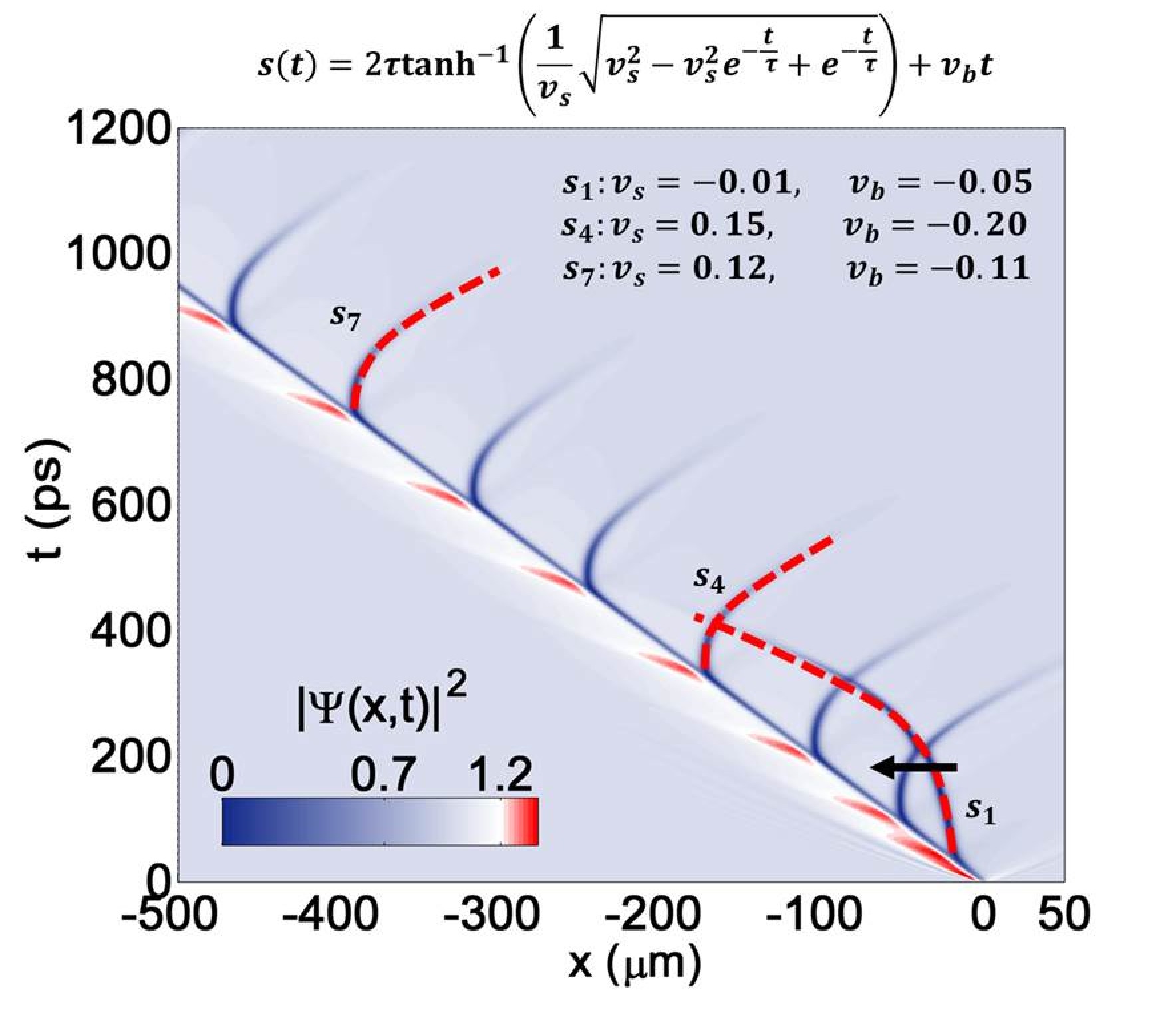}
\caption{Soliton trajectories in laboratory reference frame versus analytical predictions. Dashed lines show analytical fits to soliton trajectories. 
The first soliton \(s_1\) is created in the strongly nonequilibrium transient conditions, which results in direction of its movement which is opposite to the subsequent ones. The calculated trajectories take into account the nonzero particle current occurring in the vicinity of the defect, see  Fig.~\ref{fig:Fig_2}(f) and (\ref{eq:solitonFlow}). }
\label{fig:Fig_6}
\end{figure}
In Figure~\ref{fig:Fig_6} we present a comparison betwen the soliton trajectories calculated analytically using the formula~(\ref{33}) and numerically from Eqs.~(\ref{4})-(\ref{5}). Note that contrary to Figs.~\ref{fig:Fig_2} and~\ref{fig:Fig_3}, here the $x$ axis corresponds to the spatial variable in the laboratory reference frame. The analytical dependence was amended by the addition of the background velocity of the condensate flow, $v_b$, which is kept constant for each trajectory, and corresponds to the background density current visible in Fig.~\ref{fig:Fig_3}(f). The agreement between analytical and numerical trajectories is very good. The solitons experience gradual acceleration, which leads to their disappearance after a certain time of evolution, a phenomenon known in the studies of dark solitons in the complex Ginzburg-Landau equations~\cite{Bekki_Holes,Popp_FromDarkSolitons}. We note that the first soliton emitted at the initial transient has a velocity that is opposite to the subsequent ones (with respect to the background condensate velocity), which leads to opposite trajectory bending, as predicted by the analytical model.


\section{\label{sec:Bekki}Tanh ansatz and Bekki-Nozaki holes}

In this paper, as well as in the previous work~\cite{Smirnov_1}, it was shown that solitons of an approximate tanh shape decay in a nonequilibrium condensate due to the acceleration instability. However, the consistency of the tanh Ansatz itself may be questioned, as it assumes that the acceleration leads to the change of the phase of the condensate arbitrarily far from the soliton, through the change of parameters $A$ and $B$. This implies infinitely long-range interaction, which is inconsistent with basic physical principles.

This apparent paradox becomes even more significant in the view of the existence of the well-known Bekki-Nozaki hole solutions of the Complex Ginzburg-Landau equation (CGLE)~\cite{Bekki_Holes}. These solutions have a very similar analytic form to dark solitons. However, the crucial difference is that they possess infinite tails which are characterized by a constant density and a constant phase gradient, which corresponds to the outgoing flow of particle density. These solutions are stable in a pure CGLE, while may become unstable due to accelerating instability in a CGLE perturbed by higher order terms~\cite{Popp_HoleStability}. In contrast, the tanh-shaped solutions considered here are unstable even in the pure CGLE. 

The tanh Ansatz may be considered as a Bekki-Nozaki hole with wrong tails. The dip in the density at the soliton center is a source of particle density, which must propagate outwards. The absence of flow in the tails of the tanh waveform leads to the accumulation of density around the soliton, especially in front of it (in the direction it moves), which creates an effective potential hill. The dark soliton, being a particle with an negative effective mass, climbs up this hill, which leads to acceleration.

At a first sight, Bekki-Nozaki holes are exact and stable solutions, while tanh shaped waveforms are not stable nor exact, and they suffer from the infinite-long range interaction problem, which should rule out their physical relevance. Yet, it is striking that it is the tanh Ansatz which describes correctly the numerically observed acceleration, and not the Bekki-Nozaki stable behavior. It appears that in simulations these imperfect dark solitons are preferably created by the defect, while Bekki-Nozaki holes are very difficult to find in this system. 

The explanation of this paradox lies in the difficulty of spontaneous appearance of a waveform with "infinite" (or long enough) tails in which a phase is prepared according to the Bekki-Nozaki prescription. It is much more likely that a waveform similar to the tanh Ansatz will appear, with a flat phase in the (finite) tails. On the other hand, the issue of long-range interactions may be unimportant in practice, as the phase at long ranges may be unfolded by a slow rotation of the phase, with almost no energy cost due to the existence of low-energy Goldstone modes of phase twists~\cite{Wouters_1}.


\section{\label{sec:CONCLUSION}CONCLUSION}

In conclusion, we investigated theoretically the creation of dark soliton trains in a nonresonantly pumped exciton-polariton condensate by a moving off-resonant laser beam. We found that the frequency of soliton emission depends on the parameters of the system, such as pumping power, which optimally should be chosen close to the condensation threshold. To the contrary, it is not possible to observe stable solitons in this regime with a setup where condensate is flowing past a stationary defect, as the condensate itself becomes unstable in this configuration. The emission of solitons was related to the oscillations of the drag force. We also derived analytical formulas for the soliton trajectories using a variational approximation, obtaining very good agreement with simulations and previous results of perturbation theory.


\acknowledgments
We acknowledge support from the National Science Center grants 2015/17/B/ST3/02273, 2016/22/E/ST3/00045 and 2016/23/N/ST3/01350.

\bibliography{references_PRB}

\begin{thebibliography}{65}%
\makeatletter
\providecommand \@ifxundefined [1]{%
 \@ifx{#1\undefined}
}%
\providecommand \@ifnum [1]{%
 \ifnum #1\expandafter \@firstoftwo
 \else \expandafter \@secondoftwo
 \fi
}%
\providecommand \@ifx [1]{%
 \ifx #1\expandafter \@firstoftwo
 \else \expandafter \@secondoftwo
 \fi
}%
\providecommand \natexlab [1]{#1}%
\providecommand \enquote  [1]{``#1''}%
\providecommand \bibnamefont  [1]{#1}%
\providecommand \bibfnamefont [1]{#1}%
\providecommand \citenamefont [1]{#1}%
\providecommand \href@noop [0]{\@secondoftwo}%
\providecommand \href [0]{\begingroup \@sanitize@url \@href}%
\providecommand \@href[1]{\@@startlink{#1}\@@href}%
\providecommand \@@href[1]{\endgroup#1\@@endlink}%
\providecommand \@sanitize@url [0]{\catcode `\\12\catcode `\$12\catcode
  `\&12\catcode `\#12\catcode `\^12\catcode `\_12\catcode `\%12\relax}%
\providecommand \@@startlink[1]{}%
\providecommand \@@endlink[0]{}%
\providecommand \url  [0]{\begingroup\@sanitize@url \@url }%
\providecommand \@url [1]{\endgroup\@href {#1}{\urlprefix }}%
\providecommand \urlprefix  [0]{URL }%
\providecommand \Eprint [0]{\href }%
\providecommand \doibase [0]{http://dx.doi.org/}%
\providecommand \selectlanguage [0]{\@gobble}%
\providecommand \bibinfo  [0]{\@secondoftwo}%
\providecommand \bibfield  [0]{\@secondoftwo}%
\providecommand \translation [1]{[#1]}%
\providecommand \BibitemOpen [0]{}%
\providecommand \bibitemStop [0]{}%
\providecommand \bibitemNoStop [0]{.\EOS\space}%
\providecommand \EOS [0]{\spacefactor3000\relax}%
\providecommand \BibitemShut  [1]{\csname bibitem#1\endcsname}%
\let\auto@bib@innerbib\@empty
\bibitem [{\citenamefont {Hopfield}(1958)}]{Hopfield_1}%
  \BibitemOpen
  \bibfield  {author} {\bibinfo {author} {\bibfnamefont {J.~J.}\ \bibnamefont
  {Hopfield}},\ }\bibfield  {title} {\enquote {\bibinfo {title} {Theory of the
  contribution of excitons to the complex dielectric constant of crystals},}\
  }\href {\doibase 10.1103/PhysRev.112.1555} {\bibfield  {journal} {\bibinfo
  {journal} {Phys. Rev.}\ }\textbf {\bibinfo {volume} {112}},\ \bibinfo {pages}
  {1555--1567} (\bibinfo {year} {1958})}\BibitemShut {NoStop}%
\bibitem [{\citenamefont {Weisbuch}\ \emph {et~al.}(1992)\citenamefont
  {Weisbuch}, \citenamefont {Nishioka}, \citenamefont {Ishikawa},\ and\
  \citenamefont {Arakawa}}]{Weisbuch_1}%
  \BibitemOpen
  \bibfield  {author} {\bibinfo {author} {\bibfnamefont {C.}~\bibnamefont
  {Weisbuch}}, \bibinfo {author} {\bibfnamefont {M.}~\bibnamefont {Nishioka}},
  \bibinfo {author} {\bibfnamefont {A.}~\bibnamefont {Ishikawa}}, \ and\
  \bibinfo {author} {\bibfnamefont {Y.}~\bibnamefont {Arakawa}},\ }\bibfield
  {title} {\enquote {\bibinfo {title} {Observation of the coupled
  exciton-photon mode splitting in a semiconductor quantum microcavity},}\
  }\href {\doibase 10.1103/PhysRevLett.69.3314} {\bibfield  {journal} {\bibinfo
   {journal} {Phys. Rev. Lett.}\ }\textbf {\bibinfo {volume} {69}},\ \bibinfo
  {pages} {3314--3317} (\bibinfo {year} {1992})}\BibitemShut {NoStop}%
\bibitem [{\citenamefont {Kavokin}\ \emph {et~al.}(2008)\citenamefont
  {Kavokin}, \citenamefont {Baumberg}, \citenamefont {Malpuech},\ and\
  \citenamefont {Laussy}}]{Kavokin_1}%
  \BibitemOpen
  \bibfield  {author} {\bibinfo {author} {\bibfnamefont {Alexey}\ \bibnamefont
  {Kavokin}}, \bibinfo {author} {\bibfnamefont {Jeremy~J.}\ \bibnamefont
  {Baumberg}}, \bibinfo {author} {\bibfnamefont {Guillaume}\ \bibnamefont
  {Malpuech}}, \ and\ \bibinfo {author} {\bibfnamefont {Fabrice~P.}\
  \bibnamefont {Laussy}},\ }\href@noop {} {\emph {\bibinfo {title}
  {Microcavities}}}\ (\bibinfo {year} {2008})\BibitemShut {NoStop}%
\bibitem [{\citenamefont {Carusotto}\ and\ \citenamefont
  {Ciuti}(2013)}]{Carusotto_1}%
  \BibitemOpen
  \bibfield  {author} {\bibinfo {author} {\bibfnamefont {Iacopo}\ \bibnamefont
  {Carusotto}}\ and\ \bibinfo {author} {\bibfnamefont {Cristiano}\ \bibnamefont
  {Ciuti}},\ }\bibfield  {title} {\enquote {\bibinfo {title} {Quantum fluids of
  light},}\ }\href {\doibase 10.1103/RevModPhys.85.299} {\bibfield  {journal}
  {\bibinfo  {journal} {Rev. Mod. Phys.}\ }\textbf {\bibinfo {volume} {85}},\
  \bibinfo {pages} {299--366} (\bibinfo {year} {2013})}\BibitemShut {NoStop}%
\bibitem [{\citenamefont {Deng}\ \emph {et~al.}(2010)\citenamefont {Deng},
  \citenamefont {Haug},\ and\ \citenamefont {Yamamoto}}]{Deng_1}%
  \BibitemOpen
  \bibfield  {author} {\bibinfo {author} {\bibfnamefont {Hui}\ \bibnamefont
  {Deng}}, \bibinfo {author} {\bibfnamefont {Hartmut}\ \bibnamefont {Haug}}, \
  and\ \bibinfo {author} {\bibfnamefont {Yoshihisa}\ \bibnamefont {Yamamoto}},\
  }\bibfield  {title} {\enquote {\bibinfo {title} {Exciton-polariton
  bose-einstein condensation},}\ }\href {\doibase 10.1103/RevModPhys.82.1489}
  {\bibfield  {journal} {\bibinfo  {journal} {Rev. Mod. Phys.}\ }\textbf
  {\bibinfo {volume} {82}},\ \bibinfo {pages} {1489--1537} (\bibinfo {year}
  {2010})}\BibitemShut {NoStop}%
\bibitem [{\citenamefont {Christopoulos}\ \emph {et~al.}(2007)\citenamefont
  {Christopoulos}, \citenamefont {von H\"ogersthal}, \citenamefont {Grundy},
  \citenamefont {Lagoudakis}, \citenamefont {Kavokin}, \citenamefont
  {Baumberg}, \citenamefont {Christmann}, \citenamefont {Butt\'e},
  \citenamefont {Feltin}, \citenamefont {Carlin},\ and\ \citenamefont
  {Grandjean}}]{Christopoulos_1}%
  \BibitemOpen
  \bibfield  {author} {\bibinfo {author} {\bibfnamefont {S.}~\bibnamefont
  {Christopoulos}}, \bibinfo {author} {\bibfnamefont {G.~Baldassarri~H\"oger}\
  \bibnamefont {von H\"ogersthal}}, \bibinfo {author} {\bibfnamefont
  {A.~J.~D.}\ \bibnamefont {Grundy}}, \bibinfo {author} {\bibfnamefont {P.~G.}\
  \bibnamefont {Lagoudakis}}, \bibinfo {author} {\bibfnamefont {A.~V.}\
  \bibnamefont {Kavokin}}, \bibinfo {author} {\bibfnamefont {J.~J.}\
  \bibnamefont {Baumberg}}, \bibinfo {author} {\bibfnamefont {G.}~\bibnamefont
  {Christmann}}, \bibinfo {author} {\bibfnamefont {R.}~\bibnamefont {Butt\'e}},
  \bibinfo {author} {\bibfnamefont {E.}~\bibnamefont {Feltin}}, \bibinfo
  {author} {\bibfnamefont {J.-F.}\ \bibnamefont {Carlin}}, \ and\ \bibinfo
  {author} {\bibfnamefont {N.}~\bibnamefont {Grandjean}},\ }\bibfield  {title}
  {\enquote {\bibinfo {title} {Room-temperature polariton lasing in
  semiconductor microcavities},}\ }\href {\doibase
  10.1103/PhysRevLett.98.126405} {\bibfield  {journal} {\bibinfo  {journal}
  {Phys. Rev. Lett.}\ }\textbf {\bibinfo {volume} {98}},\ \bibinfo {pages}
  {126405} (\bibinfo {year} {2007})}\BibitemShut {NoStop}%
\bibitem [{\citenamefont {Kasprzak}\ \emph {et~al.}(2006)\citenamefont
  {Kasprzak}, \citenamefont {Richard}, \citenamefont {Kundermann},
  \citenamefont {Baas}, \citenamefont {Jeambrun}, \citenamefont {Keeling},
  \citenamefont {Marchetti}, \citenamefont {Szymanska}, \citenamefont {Andre},
  \citenamefont {Staehli}, \citenamefont {Savona}, \citenamefont {Littlewood},
  \citenamefont {Deveaud},\ and\ \citenamefont {Dang}}]{Kasprzak_1}%
  \BibitemOpen
  \bibfield  {author} {\bibinfo {author} {\bibfnamefont {J.}~\bibnamefont
  {Kasprzak}}, \bibinfo {author} {\bibfnamefont {M.}~\bibnamefont {Richard}},
  \bibinfo {author} {\bibfnamefont {S.}~\bibnamefont {Kundermann}}, \bibinfo
  {author} {\bibfnamefont {A.}~\bibnamefont {Baas}}, \bibinfo {author}
  {\bibfnamefont {P.}~\bibnamefont {Jeambrun}}, \bibinfo {author}
  {\bibfnamefont {J.~M.~J.}\ \bibnamefont {Keeling}}, \bibinfo {author}
  {\bibfnamefont {F.~M.}\ \bibnamefont {Marchetti}}, \bibinfo {author}
  {\bibfnamefont {M.~H.}\ \bibnamefont {Szymanska}}, \bibinfo {author}
  {\bibfnamefont {R.}~\bibnamefont {Andre}}, \bibinfo {author} {\bibfnamefont
  {J.~L.}\ \bibnamefont {Staehli}}, \bibinfo {author} {\bibfnamefont
  {V.}~\bibnamefont {Savona}}, \bibinfo {author} {\bibfnamefont {P.~B.}\
  \bibnamefont {Littlewood}}, \bibinfo {author} {\bibfnamefont
  {B.}~\bibnamefont {Deveaud}}, \ and\ \bibinfo {author} {\bibfnamefont
  {Le~Si}\ \bibnamefont {Dang}},\ }\bibfield  {title} {\enquote {\bibinfo
  {title} {Bose-einstein condensation of exciton polaritons},}\ }\href
  {\doibase 10.1038/nature05131} {\bibfield  {journal} {\bibinfo  {journal}
  {Nature}\ }\textbf {\bibinfo {volume} {443}},\ \bibinfo {pages} {409--414}
  (\bibinfo {year} {2006})}\BibitemShut {NoStop}%
\bibitem [{\citenamefont {K{\'e}na-Cohen}\ and\ \citenamefont
  {Forrest}(2010)}]{Cohen_1}%
  \BibitemOpen
  \bibfield  {author} {\bibinfo {author} {\bibfnamefont {S.}~\bibnamefont
  {K{\'e}na-Cohen}}\ and\ \bibinfo {author} {\bibfnamefont {S.~R.}\
  \bibnamefont {Forrest}},\ }\bibfield  {title} {\enquote {\bibinfo {title}
  {Room-temperature polariton lasing in an organic single-crystal
  microcavity},}\ }\href {http://dx.doi.org/10.1038/nphoton.2010.86} {\bibfield
   {journal} {\bibinfo  {journal} {Nature Photonics}\ }\textbf {\bibinfo
  {volume} {4}},\ \bibinfo {pages} {371--375} (\bibinfo {year}
  {2010})}\BibitemShut {NoStop}%
\bibitem [{\citenamefont {Guillet}\ and\ \citenamefont
  {Brimont}(2016)}]{Guillet_1}%
  \BibitemOpen
  \bibfield  {author} {\bibinfo {author} {\bibfnamefont {Thierry}\ \bibnamefont
  {Guillet}}\ and\ \bibinfo {author} {\bibfnamefont {Christelle}\ \bibnamefont
  {Brimont}},\ }\bibfield  {title} {\enquote {\bibinfo {title} {Polariton
  condensates at room temperature},}\ }\href {\doibase
  https://doi.org/10.1016/j.crhy.2016.07.002} {\bibfield  {journal} {\bibinfo
  {journal} {Comptes Rendus Physique}\ }\textbf {\bibinfo {volume} {17}},\
  \bibinfo {pages} {946 -- 956} (\bibinfo {year} {2016})}\BibitemShut {NoStop}%
\bibitem [{\citenamefont {Li}\ \emph {et~al.}(2013)\citenamefont {Li},
  \citenamefont {Orosz}, \citenamefont {Kamoun}, \citenamefont {Bouchoule},
  \citenamefont {Brimont}, \citenamefont {Disseix}, \citenamefont {Guillet},
  \citenamefont {Lafosse}, \citenamefont {Leroux}, \citenamefont {Leymarie},
  \citenamefont {Mexis}, \citenamefont {Mihailovic}, \citenamefont
  {Patriarche}, \citenamefont {R\'everet}, \citenamefont {Solnyshkov},
  \citenamefont {Zuniga-Perez},\ and\ \citenamefont {Malpuech}}]{Li_1}%
  \BibitemOpen
  \bibfield  {author} {\bibinfo {author} {\bibfnamefont {Feng}\ \bibnamefont
  {Li}}, \bibinfo {author} {\bibfnamefont {L.}~\bibnamefont {Orosz}}, \bibinfo
  {author} {\bibfnamefont {O.}~\bibnamefont {Kamoun}}, \bibinfo {author}
  {\bibfnamefont {S.}~\bibnamefont {Bouchoule}}, \bibinfo {author}
  {\bibfnamefont {C.}~\bibnamefont {Brimont}}, \bibinfo {author} {\bibfnamefont
  {P.}~\bibnamefont {Disseix}}, \bibinfo {author} {\bibfnamefont
  {T.}~\bibnamefont {Guillet}}, \bibinfo {author} {\bibfnamefont
  {X.}~\bibnamefont {Lafosse}}, \bibinfo {author} {\bibfnamefont
  {M.}~\bibnamefont {Leroux}}, \bibinfo {author} {\bibfnamefont
  {J.}~\bibnamefont {Leymarie}}, \bibinfo {author} {\bibfnamefont
  {M.}~\bibnamefont {Mexis}}, \bibinfo {author} {\bibfnamefont
  {M.}~\bibnamefont {Mihailovic}}, \bibinfo {author} {\bibfnamefont
  {G.}~\bibnamefont {Patriarche}}, \bibinfo {author} {\bibfnamefont
  {F.}~\bibnamefont {R\'everet}}, \bibinfo {author} {\bibfnamefont
  {D.}~\bibnamefont {Solnyshkov}}, \bibinfo {author} {\bibfnamefont
  {J.}~\bibnamefont {Zuniga-Perez}}, \ and\ \bibinfo {author} {\bibfnamefont
  {G.}~\bibnamefont {Malpuech}},\ }\bibfield  {title} {\enquote {\bibinfo
  {title} {From excitonic to photonic polariton condensate in a zno-based
  microcavity},}\ }\href {\doibase 10.1103/PhysRevLett.110.196406} {\bibfield
  {journal} {\bibinfo  {journal} {Phys. Rev. Lett.}\ }\textbf {\bibinfo
  {volume} {110}},\ \bibinfo {pages} {196406} (\bibinfo {year}
  {2013})}\BibitemShut {NoStop}%
\bibitem [{\citenamefont {Amo}\ \emph {et~al.}(2009)\citenamefont {Amo},
  \citenamefont {Lefr{\`e}re}, \citenamefont {Pigeon}, \citenamefont {Adrados},
  \citenamefont {Ciuti}, \citenamefont {Carusotto}, \citenamefont {Houdr{\'e}},
  \citenamefont {Giacobino},\ and\ \citenamefont {Bramati}}]{Amo_1}%
  \BibitemOpen
  \bibfield  {author} {\bibinfo {author} {\bibfnamefont {Alberto}\ \bibnamefont
  {Amo}}, \bibinfo {author} {\bibfnamefont {J{\'e}r{\^o}me}\ \bibnamefont
  {Lefr{\`e}re}}, \bibinfo {author} {\bibfnamefont {Simon}\ \bibnamefont
  {Pigeon}}, \bibinfo {author} {\bibfnamefont {Claire}\ \bibnamefont
  {Adrados}}, \bibinfo {author} {\bibfnamefont {Cristiano}\ \bibnamefont
  {Ciuti}}, \bibinfo {author} {\bibfnamefont {Iacopo}\ \bibnamefont
  {Carusotto}}, \bibinfo {author} {\bibfnamefont {Romuald}\ \bibnamefont
  {Houdr{\'e}}}, \bibinfo {author} {\bibfnamefont {Elisabeth}\ \bibnamefont
  {Giacobino}}, \ and\ \bibinfo {author} {\bibfnamefont {Alberto}\ \bibnamefont
  {Bramati}},\ }\bibfield  {title} {\enquote {\bibinfo {title} {Superfluidity
  of polaritons in semiconductor microcavities},}\ }\href
  {http://dx.doi.org/10.1038/nphys1364} {\bibfield  {journal} {\bibinfo
  {journal} {Nature Physics}\ }\textbf {\bibinfo {volume} {5}},\ \bibinfo
  {pages} {805} (\bibinfo {year} {2009})}\BibitemShut {NoStop}%
\bibitem [{\citenamefont {Lerario}\ \emph {et~al.}(2017)\citenamefont
  {Lerario}, \citenamefont {Fieramosca}, \citenamefont {Barachati},
  \citenamefont {Ballarini}, \citenamefont {Daskalakis}, \citenamefont
  {Dominici}, \citenamefont {De~Giorgi}, \citenamefont {Maier}, \citenamefont
  {Gigli}, \citenamefont {K{\'e}na-Cohen},\ and\ \citenamefont
  {Sanvitto}}]{Lerario_1}%
  \BibitemOpen
  \bibfield  {author} {\bibinfo {author} {\bibfnamefont {Giovanni}\
  \bibnamefont {Lerario}}, \bibinfo {author} {\bibfnamefont {Antonio}\
  \bibnamefont {Fieramosca}}, \bibinfo {author} {\bibfnamefont {F{\'a}bio}\
  \bibnamefont {Barachati}}, \bibinfo {author} {\bibfnamefont {Dario}\
  \bibnamefont {Ballarini}}, \bibinfo {author} {\bibfnamefont
  {Konstantinos~S.}\ \bibnamefont {Daskalakis}}, \bibinfo {author}
  {\bibfnamefont {Lorenzo}\ \bibnamefont {Dominici}}, \bibinfo {author}
  {\bibfnamefont {Milena}\ \bibnamefont {De~Giorgi}}, \bibinfo {author}
  {\bibfnamefont {Stefan~A.}\ \bibnamefont {Maier}}, \bibinfo {author}
  {\bibfnamefont {Giuseppe}\ \bibnamefont {Gigli}}, \bibinfo {author}
  {\bibfnamefont {St{\'e}phane}\ \bibnamefont {K{\'e}na-Cohen}}, \ and\
  \bibinfo {author} {\bibfnamefont {Daniele}\ \bibnamefont {Sanvitto}},\
  }\bibfield  {title} {\enquote {\bibinfo {title} {Room-temperature
  superfluidity in a polariton condensate},}\ }\href@noop {} {\bibfield
  {journal} {\bibinfo  {journal} {Nature Physics}\ }\textbf {\bibinfo {volume}
  {13}},\ \bibinfo {pages} {837} (\bibinfo {year} {2017})}\BibitemShut
  {NoStop}%
\bibitem [{\citenamefont {Ostrovskaya}\ \emph {et~al.}(2012)\citenamefont
  {Ostrovskaya}, \citenamefont {Abdullaev}, \citenamefont {Desyatnikov},
  \citenamefont {Fraser},\ and\ \citenamefont {Kivshar}}]{Ostrovskaya_1}%
  \BibitemOpen
  \bibfield  {author} {\bibinfo {author} {\bibfnamefont {Elena~A.}\
  \bibnamefont {Ostrovskaya}}, \bibinfo {author} {\bibfnamefont {Jasur}\
  \bibnamefont {Abdullaev}}, \bibinfo {author} {\bibfnamefont {Anton~S.}\
  \bibnamefont {Desyatnikov}}, \bibinfo {author} {\bibfnamefont {Michael~D.}\
  \bibnamefont {Fraser}}, \ and\ \bibinfo {author} {\bibfnamefont {Yuri~S.}\
  \bibnamefont {Kivshar}},\ }\bibfield  {title} {\enquote {\bibinfo {title}
  {Dissipative solitons and vortices in polariton bose-einstein condensates},}\
  }\href {\doibase 10.1103/PhysRevA.86.013636} {\bibfield  {journal} {\bibinfo
  {journal} {Phys. Rev. A}\ }\textbf {\bibinfo {volume} {86}},\ \bibinfo
  {pages} {013636} (\bibinfo {year} {2012})}\BibitemShut {NoStop}%
\bibitem [{\citenamefont {Lagoudakis}\ \emph {et~al.}(2011)\citenamefont
  {Lagoudakis}, \citenamefont {Manni}, \citenamefont {Pietka}, \citenamefont
  {Wouters}, \citenamefont {Liew}, \citenamefont {Savona}, \citenamefont
  {Kavokin}, \citenamefont {Andr\'e},\ and\ \citenamefont
  {Deveaud-Pl\'edran}}]{Lagoudakis_1}%
  \BibitemOpen
  \bibfield  {author} {\bibinfo {author} {\bibfnamefont {K.~G.}\ \bibnamefont
  {Lagoudakis}}, \bibinfo {author} {\bibfnamefont {F.}~\bibnamefont {Manni}},
  \bibinfo {author} {\bibfnamefont {B.}~\bibnamefont {Pietka}}, \bibinfo
  {author} {\bibfnamefont {M.}~\bibnamefont {Wouters}}, \bibinfo {author}
  {\bibfnamefont {T.~C.~H.}\ \bibnamefont {Liew}}, \bibinfo {author}
  {\bibfnamefont {V.}~\bibnamefont {Savona}}, \bibinfo {author} {\bibfnamefont
  {A.~V.}\ \bibnamefont {Kavokin}}, \bibinfo {author} {\bibfnamefont
  {R.}~\bibnamefont {Andr\'e}}, \ and\ \bibinfo {author} {\bibfnamefont
  {B.}~\bibnamefont {Deveaud-Pl\'edran}},\ }\bibfield  {title} {\enquote
  {\bibinfo {title} {Probing the dynamics of spontaneous quantum vortices in
  polariton superfluids},}\ }\href {\doibase 10.1103/PhysRevLett.106.115301}
  {\bibfield  {journal} {\bibinfo  {journal} {Phys. Rev. Lett.}\ }\textbf
  {\bibinfo {volume} {106}},\ \bibinfo {pages} {115301} (\bibinfo {year}
  {2011})}\BibitemShut {NoStop}%
\bibitem [{\citenamefont {Sich}\ \emph {et~al.}(2011)\citenamefont {Sich},
  \citenamefont {Krizhanovskii}, \citenamefont {Skolnick}, \citenamefont
  {Gorbach}, \citenamefont {Hartley}, \citenamefont {Skryabin}, \citenamefont
  {Cerda-M{\'e}ndez}, \citenamefont {Biermann}, \citenamefont {Hey},\ and\
  \citenamefont {Santos}}]{Sich_2}%
  \BibitemOpen
  \bibfield  {author} {\bibinfo {author} {\bibfnamefont {M.}~\bibnamefont
  {Sich}}, \bibinfo {author} {\bibfnamefont {D.~N.}\ \bibnamefont
  {Krizhanovskii}}, \bibinfo {author} {\bibfnamefont {M.~S.}\ \bibnamefont
  {Skolnick}}, \bibinfo {author} {\bibfnamefont {A.~V.}\ \bibnamefont
  {Gorbach}}, \bibinfo {author} {\bibfnamefont {R.}~\bibnamefont {Hartley}},
  \bibinfo {author} {\bibfnamefont {D.~V.}\ \bibnamefont {Skryabin}}, \bibinfo
  {author} {\bibfnamefont {E.~A.}\ \bibnamefont {Cerda-M{\'e}ndez}}, \bibinfo
  {author} {\bibfnamefont {K.}~\bibnamefont {Biermann}}, \bibinfo {author}
  {\bibfnamefont {R.}~\bibnamefont {Hey}}, \ and\ \bibinfo {author}
  {\bibfnamefont {P.~V.}\ \bibnamefont {Santos}},\ }\bibfield  {title}
  {\enquote {\bibinfo {title} {Observation of bright polariton solitons in a
  semiconductor microcavity},}\ }\href
  {http://dx.doi.org/10.1038/nphoton.2011.267} {\ \textbf {\bibinfo {volume}
  {6}},\ \bibinfo {pages} {50 EP --} (\bibinfo {year} {2011})},\ \bibinfo
  {note} {article}\BibitemShut {NoStop}%
\bibitem [{\citenamefont {Solnyshkov}\ \emph {et~al.}(2017)\citenamefont
  {Solnyshkov}, \citenamefont {Bleu}, \citenamefont {Teklu},\ and\
  \citenamefont {Malpuech}}]{Malpuech_1}%
  \BibitemOpen
  \bibfield  {author} {\bibinfo {author} {\bibfnamefont {D.~D.}\ \bibnamefont
  {Solnyshkov}}, \bibinfo {author} {\bibfnamefont {O.}~\bibnamefont {Bleu}},
  \bibinfo {author} {\bibfnamefont {B.}~\bibnamefont {Teklu}}, \ and\ \bibinfo
  {author} {\bibfnamefont {G.}~\bibnamefont {Malpuech}},\ }\bibfield  {title}
  {\enquote {\bibinfo {title} {Chirality of topological gap solitons in bosonic
  dimer chains},}\ }\href {\doibase 10.1103/PhysRevLett.118.023901} {\bibfield
  {journal} {\bibinfo  {journal} {Phys. Rev. Lett.}\ }\textbf {\bibinfo
  {volume} {118}},\ \bibinfo {pages} {023901} (\bibinfo {year}
  {2017})}\BibitemShut {NoStop}%
\bibitem [{\citenamefont {Kartashov}\ and\ \citenamefont
  {Skryabin}(2016{\natexlab{a}})}]{Kartashov_2}%
  \BibitemOpen
  \bibfield  {author} {\bibinfo {author} {\bibfnamefont {Yaroslav~V.}\
  \bibnamefont {Kartashov}}\ and\ \bibinfo {author} {\bibfnamefont {Dmitry~V.}\
  \bibnamefont {Skryabin}},\ }\bibfield  {title} {\enquote {\bibinfo {title}
  {Modulational instability and solitary waves in polariton topological
  insulators},}\ }\href {\doibase 10.1364/OPTICA.3.001228} {\bibfield
  {journal} {\bibinfo  {journal} {Optica}\ }\textbf {\bibinfo {volume} {3}},\
  \bibinfo {pages} {1228--1236} (\bibinfo {year}
  {2016}{\natexlab{a}})}\BibitemShut {NoStop}%
\bibitem [{\citenamefont {Frantzeskakis}(2010)}]{Frantzeskakis_1}%
  \BibitemOpen
  \bibfield  {author} {\bibinfo {author} {\bibfnamefont {D~J}\ \bibnamefont
  {Frantzeskakis}},\ }\bibfield  {title} {\enquote {\bibinfo {title} {Dark
  solitons in atomic bose–einstein condensates: from theory to
  experiments},}\ }\href {http://stacks.iop.org/1751-8121/43/i=21/a=213001}
  {\bibfield  {journal} {\bibinfo  {journal} {Journal of Physics A:
  Mathematical and Theoretical}\ }\textbf {\bibinfo {volume} {43}},\ \bibinfo
  {pages} {213001} (\bibinfo {year} {2010})}\BibitemShut {NoStop}%
\bibitem [{\citenamefont {Kevrekidis}\ \emph {et~al.}(2015)\citenamefont
  {Kevrekidis}, \citenamefont {Frantzeskakis},\ and\ \citenamefont
  {Carretero-Gonz\'alez}}]{Kevrekidis_1}%
  \BibitemOpen
  \bibfield  {author} {\bibinfo {author} {\bibfnamefont {P.}~\bibnamefont
  {Kevrekidis}}, \bibinfo {author} {\bibfnamefont {D.}~\bibnamefont
  {Frantzeskakis}}, \ and\ \bibinfo {author} {\bibfnamefont {R.}~\bibnamefont
  {Carretero-Gonz\'alez}},\ }\href@noop {} {\emph {\bibinfo {title} {The
  Defocusing Nonlinear Schr\"odinger Equation}}}\ (\bibinfo  {publisher}
  {Society for Industrial and Applied Mathematics},\ \bibinfo {address}
  {Philadelphia, PA},\ \bibinfo {year} {2015})\BibitemShut {NoStop}%
\bibitem [{\citenamefont {Kivshar}\ and\ \citenamefont
  {Kr\'olikowski}(1995)}]{Kivshar_1}%
  \BibitemOpen
  \bibfield  {author} {\bibinfo {author} {\bibfnamefont {Yuri~S.}\ \bibnamefont
  {Kivshar}}\ and\ \bibinfo {author} {\bibfnamefont {Wiesław}\ \bibnamefont
  {Kr\'olikowski}},\ }\bibfield  {title} {\enquote {\bibinfo {title}
  {Lagrangian approach for dark solitons},}\ }\href {\doibase
  http://dx.doi.org/10.1016/0030-4018(94)00644-A} {\bibfield  {journal}
  {\bibinfo  {journal} {Optics Communications}\ }\textbf {\bibinfo {volume}
  {114}},\ \bibinfo {pages} {353 -- 362} (\bibinfo {year} {1995})}\BibitemShut
  {NoStop}%
\bibitem [{\citenamefont {Kivshar}\ and\ \citenamefont
  {Malomed}(1989)}]{Kivshar_2}%
  \BibitemOpen
  \bibfield  {author} {\bibinfo {author} {\bibfnamefont {Yuri~S.}\ \bibnamefont
  {Kivshar}}\ and\ \bibinfo {author} {\bibfnamefont {Boris~A.}\ \bibnamefont
  {Malomed}},\ }\bibfield  {title} {\enquote {\bibinfo {title} {Dynamics of
  solitons in nearly integrable systems},}\ }\href {\doibase
  10.1103/RevModPhys.61.763} {\bibfield  {journal} {\bibinfo  {journal} {Rev.
  Mod. Phys.}\ }\textbf {\bibinfo {volume} {61}},\ \bibinfo {pages} {763--915}
  (\bibinfo {year} {1989})}\BibitemShut {NoStop}%
\bibitem [{\citenamefont {Infeld}\ and\ \citenamefont
  {Rowlands}(1990)}]{Infeld_1}%
  \BibitemOpen
  \bibfield  {author} {\bibinfo {author} {\bibfnamefont {E.}~\bibnamefont
  {Infeld}}\ and\ \bibinfo {author} {\bibfnamefont {G.}~\bibnamefont
  {Rowlands}},\ }\href@noop {} {\emph {\bibinfo {title} {Nonlinear Waves,
  Solitons and Chaos}}}\ (\bibinfo {year} {1990})\BibitemShut {NoStop}%
\bibitem [{Sic(2016)}]{Sich_1}%
  \BibitemOpen
  \bibfield  {title} {\enquote {\bibinfo {title} {Soliton physics with
  semiconductor exciton–polaritons in confined systems},}\ }\href@noop {}
  {\bibfield  {journal} {\bibinfo  {journal} {Comptes Rendus Physique}\
  }\textbf {\bibinfo {volume} {17}},\ \bibinfo {pages} {908 -- 919} (\bibinfo
  {year} {2016})}\BibitemShut {NoStop}%
\bibitem [{\citenamefont {Tanese}\ \emph {et~al.}(2013)\citenamefont {Tanese},
  \citenamefont {Flayac}, \citenamefont {Solnyshkov}, \citenamefont {Amo},
  \citenamefont {Lema{\^i}tre}, \citenamefont {Galopin}, \citenamefont
  {Braive}, \citenamefont {Senellart}, \citenamefont {Sagnes}, \citenamefont
  {Malpuech},\ and\ \citenamefont {Bloch}}]{Tanese_1}%
  \BibitemOpen
  \bibfield  {author} {\bibinfo {author} {\bibfnamefont {D.}~\bibnamefont
  {Tanese}}, \bibinfo {author} {\bibfnamefont {H.}~\bibnamefont {Flayac}},
  \bibinfo {author} {\bibfnamefont {D.}~\bibnamefont {Solnyshkov}}, \bibinfo
  {author} {\bibfnamefont {A.}~\bibnamefont {Amo}}, \bibinfo {author}
  {\bibfnamefont {A.}~\bibnamefont {Lema{\^i}tre}}, \bibinfo {author}
  {\bibfnamefont {E.}~\bibnamefont {Galopin}}, \bibinfo {author} {\bibfnamefont
  {R.}~\bibnamefont {Braive}}, \bibinfo {author} {\bibfnamefont
  {P.}~\bibnamefont {Senellart}}, \bibinfo {author} {\bibfnamefont
  {I.}~\bibnamefont {Sagnes}}, \bibinfo {author} {\bibfnamefont
  {G.}~\bibnamefont {Malpuech}}, \ and\ \bibinfo {author} {\bibfnamefont
  {J.}~\bibnamefont {Bloch}},\ }\bibfield  {title} {\enquote {\bibinfo {title}
  {Polariton condensation in solitonic gap states in a one-dimensional periodic
  potential},}\ }\href {http://dx.doi.org/10.1038/ncomms2760} {\bibfield
  {journal} {\bibinfo  {journal} {Nature Communications}\ }\textbf {\bibinfo
  {volume} {4}},\ \bibinfo {pages} {1749} (\bibinfo {year} {2013})}\BibitemShut
  {NoStop}%
\bibitem [{\citenamefont {Flayac}\ \emph {et~al.}(2011)\citenamefont {Flayac},
  \citenamefont {Solnyshkov},\ and\ \citenamefont {Malpuech}}]{Flayac_1}%
  \BibitemOpen
  \bibfield  {author} {\bibinfo {author} {\bibfnamefont {H.}~\bibnamefont
  {Flayac}}, \bibinfo {author} {\bibfnamefont {D.~D.}\ \bibnamefont
  {Solnyshkov}}, \ and\ \bibinfo {author} {\bibfnamefont {G.}~\bibnamefont
  {Malpuech}},\ }\bibfield  {title} {\enquote {\bibinfo {title} {Oblique
  half-solitons and their generation in exciton-polariton condensates},}\
  }\href {\doibase 10.1103/PhysRevB.83.193305} {\bibfield  {journal} {\bibinfo
  {journal} {Phys. Rev. B}\ }\textbf {\bibinfo {volume} {83}},\ \bibinfo
  {pages} {193305} (\bibinfo {year} {2011})}\BibitemShut {NoStop}%
\bibitem [{\citenamefont {Hivet}\ \emph {et~al.}(2012)\citenamefont {Hivet},
  \citenamefont {Flayac}, \citenamefont {Solnyshkov}, \citenamefont {Tanese},
  \citenamefont {Boulier}, \citenamefont {Andreoli}, \citenamefont {Giacobino},
  \citenamefont {Bloch}, \citenamefont {Bramati}, \citenamefont {Malpuech},\
  and\ \citenamefont {Amo}}]{Hivet_1}%
  \BibitemOpen
  \bibfield  {author} {\bibinfo {author} {\bibfnamefont {R.}~\bibnamefont
  {Hivet}}, \bibinfo {author} {\bibfnamefont {H.}~\bibnamefont {Flayac}},
  \bibinfo {author} {\bibfnamefont {D.~D.}\ \bibnamefont {Solnyshkov}},
  \bibinfo {author} {\bibfnamefont {D.}~\bibnamefont {Tanese}}, \bibinfo
  {author} {\bibfnamefont {T.}~\bibnamefont {Boulier}}, \bibinfo {author}
  {\bibfnamefont {D.}~\bibnamefont {Andreoli}}, \bibinfo {author}
  {\bibfnamefont {E.}~\bibnamefont {Giacobino}}, \bibinfo {author}
  {\bibfnamefont {J.}~\bibnamefont {Bloch}}, \bibinfo {author} {\bibfnamefont
  {A.}~\bibnamefont {Bramati}}, \bibinfo {author} {\bibfnamefont
  {G.}~\bibnamefont {Malpuech}}, \ and\ \bibinfo {author} {\bibfnamefont
  {A.}~\bibnamefont {Amo}},\ }\bibfield  {title} {\enquote {\bibinfo {title}
  {Half-solitons in a polariton quantum fluid behave like magnetic
  monopoles},}\ }\href {http://dx.doi.org/10.1038/nphys2406} {\bibfield
  {journal} {\bibinfo  {journal} {Nature Physics}\ }\textbf {\bibinfo {volume}
  {8}},\ \bibinfo {pages} {724--728} (\bibinfo {year} {2012})}\BibitemShut
  {NoStop}%
\bibitem [{\citenamefont {Smirnov}\ \emph {et~al.}(2014)\citenamefont
  {Smirnov}, \citenamefont {Smirnova}, \citenamefont {Ostrovskaya},\ and\
  \citenamefont {Kivshar}}]{Smirnov_1}%
  \BibitemOpen
  \bibfield  {author} {\bibinfo {author} {\bibfnamefont {Lev~A.}\ \bibnamefont
  {Smirnov}}, \bibinfo {author} {\bibfnamefont {Daria~A.}\ \bibnamefont
  {Smirnova}}, \bibinfo {author} {\bibfnamefont {Elena~A.}\ \bibnamefont
  {Ostrovskaya}}, \ and\ \bibinfo {author} {\bibfnamefont {Yuri~S.}\
  \bibnamefont {Kivshar}},\ }\bibfield  {title} {\enquote {\bibinfo {title}
  {Dynamics and stability of dark solitons in exciton-polariton condensates},}\
  }\href {\doibase 10.1103/PhysRevB.89.235310} {\bibfield  {journal} {\bibinfo
  {journal} {Phys. Rev. B}\ }\textbf {\bibinfo {volume} {89}},\ \bibinfo
  {pages} {235310} (\bibinfo {year} {2014})}\BibitemShut {NoStop}%
\bibitem [{\citenamefont {Pinsker}\ and\ \citenamefont
  {Flayac}(2014)}]{Pinsker_1}%
  \BibitemOpen
  \bibfield  {author} {\bibinfo {author} {\bibfnamefont {F.}~\bibnamefont
  {Pinsker}}\ and\ \bibinfo {author} {\bibfnamefont {H.}~\bibnamefont
  {Flayac}},\ }\bibfield  {title} {\enquote {\bibinfo {title} {On-demand dark
  soliton train manipulation in a spinor polariton condensate},}\ }\href
  {\doibase 10.1103/PhysRevLett.112.140405} {\bibfield  {journal} {\bibinfo
  {journal} {Phys. Rev. Lett.}\ }\textbf {\bibinfo {volume} {112}},\ \bibinfo
  {pages} {140405} (\bibinfo {year} {2014})}\BibitemShut {NoStop}%
\bibitem [{\citenamefont {Pinsker}\ and\ \citenamefont
  {Berloff}(2014)}]{Pinsker_2}%
  \BibitemOpen
  \bibfield  {author} {\bibinfo {author} {\bibfnamefont {Florian}\ \bibnamefont
  {Pinsker}}\ and\ \bibinfo {author} {\bibfnamefont {Natalia~G.}\ \bibnamefont
  {Berloff}},\ }\bibfield  {title} {\enquote {\bibinfo {title} {Transitions and
  excitations in a superfluid stream passing small impurities},}\ }\href
  {\doibase 10.1103/PhysRevA.89.053605} {\bibfield  {journal} {\bibinfo
  {journal} {Phys. Rev. A}\ }\textbf {\bibinfo {volume} {89}},\ \bibinfo
  {pages} {053605} (\bibinfo {year} {2014})}\BibitemShut {NoStop}%
\bibitem [{\citenamefont {Pinsker}\ and\ \citenamefont
  {Flayac}(2016)}]{Pinsker_3}%
  \BibitemOpen
  \bibfield  {author} {\bibinfo {author} {\bibfnamefont {F.}~\bibnamefont
  {Pinsker}}\ and\ \bibinfo {author} {\bibfnamefont {H.}~\bibnamefont
  {Flayac}},\ }\bibfield  {title} {\enquote {\bibinfo {title} {Bright solitons
  in non-equilibrium coherent quantum matter},}\ }\href@noop {} {\ \textbf
  {\bibinfo {volume} {472}} (\bibinfo {year} {2016})}\BibitemShut {NoStop}%
\bibitem [{\citenamefont {Xue}\ and\ \citenamefont
  {Matuszewski}(2014)}]{Xue_1}%
  \BibitemOpen
  \bibfield  {author} {\bibinfo {author} {\bibfnamefont {Yan}\ \bibnamefont
  {Xue}}\ and\ \bibinfo {author} {\bibfnamefont {Micha\l{}}\ \bibnamefont
  {Matuszewski}},\ }\bibfield  {title} {\enquote {\bibinfo {title} {Creation
  and abrupt decay of a quasistationary dark soliton in a polariton
  condensate},}\ }\href {\doibase 10.1103/PhysRevLett.112.216401} {\bibfield
  {journal} {\bibinfo  {journal} {Phys. Rev. Lett.}\ }\textbf {\bibinfo
  {volume} {112}},\ \bibinfo {pages} {216401} (\bibinfo {year}
  {2014})}\BibitemShut {NoStop}%
\bibitem [{\citenamefont {Chen}\ \emph {et~al.}(2015)\citenamefont {Chen},
  \citenamefont {Hsieh},\ and\ \citenamefont {Cheng}}]{Chen_1}%
  \BibitemOpen
  \bibfield  {author} {\bibinfo {author} {\bibfnamefont {Ting-Wei}\
  \bibnamefont {Chen}}, \bibinfo {author} {\bibfnamefont {Wen-Feng}\
  \bibnamefont {Hsieh}}, \ and\ \bibinfo {author} {\bibfnamefont {Szu-Cheng}\
  \bibnamefont {Cheng}},\ }\bibfield  {title} {\enquote {\bibinfo {title}
  {Stable gray soliton pinned by a defect in a microcavity-polariton
  condensate},}\ }\href {\doibase 10.1364/OE.23.024974} {\bibfield  {journal}
  {\bibinfo  {journal} {Opt. Express}\ }\textbf {\bibinfo {volume} {23}},\
  \bibinfo {pages} {24974--24983} (\bibinfo {year} {2015})}\BibitemShut
  {NoStop}%
\bibitem [{\citenamefont {Kartashov}\ and\ \citenamefont
  {Skryabin}(2016{\natexlab{b}})}]{Kartashov_1}%
  \BibitemOpen
  \bibfield  {author} {\bibinfo {author} {\bibfnamefont {Yaroslav~V.}\
  \bibnamefont {Kartashov}}\ and\ \bibinfo {author} {\bibfnamefont {Dmitry~V.}\
  \bibnamefont {Skryabin}},\ }\bibfield  {title} {\enquote {\bibinfo {title}
  {Temporal dark polariton solitons},}\ }\href {\doibase 10.1364/OL.41.001760}
  {\bibfield  {journal} {\bibinfo  {journal} {Opt. Lett.}\ }\textbf {\bibinfo
  {volume} {41}},\ \bibinfo {pages} {1760--1763} (\bibinfo {year}
  {2016}{\natexlab{b}})}\BibitemShut {NoStop}%
\bibitem [{\citenamefont {Pigeon}\ \emph {et~al.}(2011)\citenamefont {Pigeon},
  \citenamefont {Carusotto},\ and\ \citenamefont {Ciuti}}]{Piegon_1}%
  \BibitemOpen
  \bibfield  {author} {\bibinfo {author} {\bibfnamefont {S.}~\bibnamefont
  {Pigeon}}, \bibinfo {author} {\bibfnamefont {I.}~\bibnamefont {Carusotto}}, \
  and\ \bibinfo {author} {\bibfnamefont {C.}~\bibnamefont {Ciuti}},\ }\bibfield
   {title} {\enquote {\bibinfo {title} {Hydrodynamic nucleation of vortices and
  solitons in a resonantly excited polariton superfluid},}\ }\href {\doibase
  10.1103/PhysRevB.83.144513} {\bibfield  {journal} {\bibinfo  {journal} {Phys.
  Rev. B}\ }\textbf {\bibinfo {volume} {83}},\ \bibinfo {pages} {144513}
  (\bibinfo {year} {2011})}\BibitemShut {NoStop}%
\bibitem [{\citenamefont {Cancellieri}\ \emph {et~al.}(2015)\citenamefont
  {Cancellieri}, \citenamefont {Chana}, \citenamefont {Sich}, \citenamefont
  {Krizhanovskii}, \citenamefont {Skolnick},\ and\ \citenamefont
  {Whittaker}}]{Cancellieri_1}%
  \BibitemOpen
  \bibfield  {author} {\bibinfo {author} {\bibfnamefont {E.}~\bibnamefont
  {Cancellieri}}, \bibinfo {author} {\bibfnamefont {J.~K.}\ \bibnamefont
  {Chana}}, \bibinfo {author} {\bibfnamefont {M.}~\bibnamefont {Sich}},
  \bibinfo {author} {\bibfnamefont {D.~N.}\ \bibnamefont {Krizhanovskii}},
  \bibinfo {author} {\bibfnamefont {M.~S.}\ \bibnamefont {Skolnick}}, \ and\
  \bibinfo {author} {\bibfnamefont {D.~M.}\ \bibnamefont {Whittaker}},\
  }\bibfield  {title} {\enquote {\bibinfo {title} {Logic gates with bright
  dissipative polariton solitons in bragg cavity systems},}\ }\href {\doibase
  10.1103/PhysRevB.92.174528} {\bibfield  {journal} {\bibinfo  {journal} {Phys.
  Rev. B}\ }\textbf {\bibinfo {volume} {92}},\ \bibinfo {pages} {174528}
  (\bibinfo {year} {2015})}\BibitemShut {NoStop}%
\bibitem [{\citenamefont {Ma}\ \emph {et~al.}(2017)\citenamefont {Ma},
  \citenamefont {Egorov},\ and\ \citenamefont {Schumacher}}]{Ma_1}%
  \BibitemOpen
  \bibfield  {author} {\bibinfo {author} {\bibfnamefont {Xuekai}\ \bibnamefont
  {Ma}}, \bibinfo {author} {\bibfnamefont {Oleg~A.}\ \bibnamefont {Egorov}}, \
  and\ \bibinfo {author} {\bibfnamefont {Stefan}\ \bibnamefont {Schumacher}},\
  }\bibfield  {title} {\enquote {\bibinfo {title} {Creation and manipulation of
  stable dark solitons and vortices in microcavity polariton condensates},}\
  }\href {\doibase 10.1103/PhysRevLett.118.157401} {\bibfield  {journal}
  {\bibinfo  {journal} {Phys. Rev. Lett.}\ }\textbf {\bibinfo {volume} {118}},\
  \bibinfo {pages} {157401} (\bibinfo {year} {2017})}\BibitemShut {NoStop}%
\bibitem [{\citenamefont {Goblot}\ \emph {et~al.}(2016)\citenamefont {Goblot},
  \citenamefont {Nguyen}, \citenamefont {Carusotto}, \citenamefont {Galopin},
  \citenamefont {Lema\^{\i}tre}, \citenamefont {Sagnes}, \citenamefont {Amo},\
  and\ \citenamefont {Bloch}}]{Goblot_1}%
  \BibitemOpen
  \bibfield  {author} {\bibinfo {author} {\bibfnamefont {V.}~\bibnamefont
  {Goblot}}, \bibinfo {author} {\bibfnamefont {H.~S.}\ \bibnamefont {Nguyen}},
  \bibinfo {author} {\bibfnamefont {I.}~\bibnamefont {Carusotto}}, \bibinfo
  {author} {\bibfnamefont {E.}~\bibnamefont {Galopin}}, \bibinfo {author}
  {\bibfnamefont {A.}~\bibnamefont {Lema\^{\i}tre}}, \bibinfo {author}
  {\bibfnamefont {I.}~\bibnamefont {Sagnes}}, \bibinfo {author} {\bibfnamefont
  {A.}~\bibnamefont {Amo}}, \ and\ \bibinfo {author} {\bibfnamefont
  {J.}~\bibnamefont {Bloch}},\ }\bibfield  {title} {\enquote {\bibinfo {title}
  {Phase-controlled bistability of a dark soliton train in a polariton
  fluid},}\ }\href {\doibase 10.1103/PhysRevLett.117.217401} {\bibfield
  {journal} {\bibinfo  {journal} {Phys. Rev. Lett.}\ }\textbf {\bibinfo
  {volume} {117}},\ \bibinfo {pages} {217401} (\bibinfo {year}
  {2016})}\BibitemShut {NoStop}%
\bibitem [{\citenamefont {Larr\'e}\ \emph {et~al.}(2012)\citenamefont
  {Larr\'e}, \citenamefont {Pavloff},\ and\ \citenamefont
  {Kamchatnov}}]{Larre_1}%
  \BibitemOpen
  \bibfield  {author} {\bibinfo {author} {\bibfnamefont {P.-\'E.}\ \bibnamefont
  {Larr\'e}}, \bibinfo {author} {\bibfnamefont {N.}~\bibnamefont {Pavloff}}, \
  and\ \bibinfo {author} {\bibfnamefont {A.~M.}\ \bibnamefont {Kamchatnov}},\
  }\bibfield  {title} {\enquote {\bibinfo {title} {Wave pattern induced by a
  localized obstacle in the flow of a one-dimensional polariton condensate},}\
  }\href {\doibase 10.1103/PhysRevB.86.165304} {\bibfield  {journal} {\bibinfo
  {journal} {Phys. Rev. B}\ }\textbf {\bibinfo {volume} {86}},\ \bibinfo
  {pages} {165304} (\bibinfo {year} {2012})}\BibitemShut {NoStop}%
\bibitem [{\citenamefont {Kamchatnov}\ \emph {et~al.}(2002)\citenamefont
  {Kamchatnov}, \citenamefont {Kraenkel},\ and\ \citenamefont
  {Umarov}}]{Kamchatnov_1}%
  \BibitemOpen
  \bibfield  {author} {\bibinfo {author} {\bibfnamefont {A.~M.}\ \bibnamefont
  {Kamchatnov}}, \bibinfo {author} {\bibfnamefont {R.~A.}\ \bibnamefont
  {Kraenkel}}, \ and\ \bibinfo {author} {\bibfnamefont {B.~A.}\ \bibnamefont
  {Umarov}},\ }\bibfield  {title} {\enquote {\bibinfo {title} {Asymptotic
  soliton train solutions of the defocusing nonlinear schr\"odinger
  equation},}\ }\href {\doibase 10.1103/PhysRevE.66.036609} {\bibfield
  {journal} {\bibinfo  {journal} {Phys. Rev. E}\ }\textbf {\bibinfo {volume}
  {66}},\ \bibinfo {pages} {036609} (\bibinfo {year} {2002})}\BibitemShut
  {NoStop}%
\bibitem [{\citenamefont {Kamchatnov}\ and\ \citenamefont
  {Kartashov}(2013)}]{Kamchatnov_2}%
  \BibitemOpen
  \bibfield  {author} {\bibinfo {author} {\bibfnamefont {A.~M.}\ \bibnamefont
  {Kamchatnov}}\ and\ \bibinfo {author} {\bibfnamefont {Y.~V.}\ \bibnamefont
  {Kartashov}},\ }\bibfield  {title} {\enquote {\bibinfo {title} {Oblique
  breathers generated by a flow of two-component bose-einstein condensates past
  a polarized obstacle},}\ }\href {\doibase 10.1103/PhysRevLett.111.140402}
  {\bibfield  {journal} {\bibinfo  {journal} {Phys. Rev. Lett.}\ }\textbf
  {\bibinfo {volume} {111}},\ \bibinfo {pages} {140402} (\bibinfo {year}
  {2013})}\BibitemShut {NoStop}%
\bibitem [{\citenamefont {Kamchatnov}\ and\ \citenamefont
  {Pavloff}(2012)}]{Kamchatnov_3}%
  \BibitemOpen
  \bibfield  {author} {\bibinfo {author} {\bibfnamefont {A.~M.}\ \bibnamefont
  {Kamchatnov}}\ and\ \bibinfo {author} {\bibfnamefont {N.}~\bibnamefont
  {Pavloff}},\ }\bibfield  {title} {\enquote {\bibinfo {title} {Generation of
  dispersive shock waves by the flow of a bose-einstein condensate past a
  narrow obstacle},}\ }\href {\doibase 10.1103/PhysRevA.85.033603} {\bibfield
  {journal} {\bibinfo  {journal} {Phys. Rev. A}\ }\textbf {\bibinfo {volume}
  {85}},\ \bibinfo {pages} {033603} (\bibinfo {year} {2012})}\BibitemShut
  {NoStop}%
\bibitem [{\citenamefont {Larr\'e}\ \emph {et~al.}(2013)\citenamefont
  {Larr\'e}, \citenamefont {Pavloff},\ and\ \citenamefont
  {Kamchatnov}}]{Larre_2}%
  \BibitemOpen
  \bibfield  {author} {\bibinfo {author} {\bibfnamefont {P.-\'E.}\ \bibnamefont
  {Larr\'e}}, \bibinfo {author} {\bibfnamefont {N.}~\bibnamefont {Pavloff}}, \
  and\ \bibinfo {author} {\bibfnamefont {A.~M.}\ \bibnamefont {Kamchatnov}},\
  }\bibfield  {title} {\enquote {\bibinfo {title} {Polarization hydrodynamics
  in a one-dimensional polariton condensate},}\ }\href {\doibase
  10.1103/PhysRevB.88.224503} {\bibfield  {journal} {\bibinfo  {journal} {Phys.
  Rev. B}\ }\textbf {\bibinfo {volume} {88}},\ \bibinfo {pages} {224503}
  (\bibinfo {year} {2013})}\BibitemShut {NoStop}%
\bibitem [{\citenamefont {Ter\c{c}as}\ and\ \citenamefont
  {Mendon\c{c}a}(2016)}]{Tercas_1}%
  \BibitemOpen
  \bibfield  {author} {\bibinfo {author} {\bibfnamefont {H.}~\bibnamefont
  {Ter\c{c}as}}\ and\ \bibinfo {author} {\bibfnamefont {J.T.}\ \bibnamefont
  {Mendon\c{c}a}},\ }\bibfield  {title} {\enquote {\bibinfo {title}
  {Exciton-polariton wakefields in semiconductor microcavities},}\ }\href@noop
  {} {\bibfield  {journal} {\bibinfo  {journal} {Physics Letters A}\ }\textbf
  {\bibinfo {volume} {380}},\ \bibinfo {pages} {822 -- 827} (\bibinfo {year}
  {2016})}\BibitemShut {NoStop}%
\bibitem [{\citenamefont {Hayat}\ \emph {et~al.}(2012)\citenamefont {Hayat},
  \citenamefont {Lange}, \citenamefont {Rozema}, \citenamefont {Darabi},
  \citenamefont {van Driel}, \citenamefont {Steinberg}, \citenamefont {Nelsen},
  \citenamefont {Snoke}, \citenamefont {Pfeiffer},\ and\ \citenamefont
  {West}}]{Hayat_1}%
  \BibitemOpen
  \bibfield  {author} {\bibinfo {author} {\bibfnamefont {Alex}\ \bibnamefont
  {Hayat}}, \bibinfo {author} {\bibfnamefont {Christoph}\ \bibnamefont
  {Lange}}, \bibinfo {author} {\bibfnamefont {Lee~A.}\ \bibnamefont {Rozema}},
  \bibinfo {author} {\bibfnamefont {Ardavan}\ \bibnamefont {Darabi}}, \bibinfo
  {author} {\bibfnamefont {Henry~M.}\ \bibnamefont {van Driel}}, \bibinfo
  {author} {\bibfnamefont {Aephraim~M.}\ \bibnamefont {Steinberg}}, \bibinfo
  {author} {\bibfnamefont {Bryan}\ \bibnamefont {Nelsen}}, \bibinfo {author}
  {\bibfnamefont {David~W.}\ \bibnamefont {Snoke}}, \bibinfo {author}
  {\bibfnamefont {Loren~N.}\ \bibnamefont {Pfeiffer}}, \ and\ \bibinfo {author}
  {\bibfnamefont {Kenneth~W.}\ \bibnamefont {West}},\ }\bibfield  {title}
  {\enquote {\bibinfo {title} {Dynamic stark effect in strongly coupled
  microcavity exciton polaritons},}\ }\href {\doibase
  10.1103/PhysRevLett.109.033605} {\bibfield  {journal} {\bibinfo  {journal}
  {Phys. Rev. Lett.}\ }\textbf {\bibinfo {volume} {109}},\ \bibinfo {pages}
  {033605} (\bibinfo {year} {2012})}\BibitemShut {NoStop}%
\bibitem [{\citenamefont {Zhang}\ \emph {et~al.}(2015)\citenamefont {Zhang},
  \citenamefont {Wu}, \citenamefont {Wang}, \citenamefont {Ma}, \citenamefont
  {Li},\ and\ \citenamefont {Rao}}]{Zhang_1}%
  \BibitemOpen
  \bibfield  {author} {\bibinfo {author} {\bibfnamefont {W.~L.}\ \bibnamefont
  {Zhang}}, \bibinfo {author} {\bibfnamefont {X.~M.}\ \bibnamefont {Wu}},
  \bibinfo {author} {\bibfnamefont {F.}~\bibnamefont {Wang}}, \bibinfo {author}
  {\bibfnamefont {R.}~\bibnamefont {Ma}}, \bibinfo {author} {\bibfnamefont
  {X.~F.}\ \bibnamefont {Li}}, \ and\ \bibinfo {author} {\bibfnamefont {Y.~J.}\
  \bibnamefont {Rao}},\ }\bibfield  {title} {\enquote {\bibinfo {title} {Stark
  effect induced microcavity polariton solitons},}\ }\href {\doibase
  10.1364/OE.23.015762} {\bibfield  {journal} {\bibinfo  {journal} {Opt.
  Express}\ }\textbf {\bibinfo {volume} {23}},\ \bibinfo {pages} {15762--15767}
  (\bibinfo {year} {2015})}\BibitemShut {NoStop}%
\bibitem [{\citenamefont {Hakim}(1997)}]{Hakim_1}%
  \BibitemOpen
  \bibfield  {author} {\bibinfo {author} {\bibfnamefont {Vincent}\ \bibnamefont
  {Hakim}},\ }\bibfield  {title} {\enquote {\bibinfo {title} {Nonlinear
  schr\"odinger flow past an obstacle in one dimension},}\ }\href {\doibase
  10.1103/PhysRevE.55.2835} {\bibfield  {journal} {\bibinfo  {journal} {Phys.
  Rev. E}\ }\textbf {\bibinfo {volume} {55}},\ \bibinfo {pages} {2835--2845}
  (\bibinfo {year} {1997})}\BibitemShut {NoStop}%
\bibitem [{\citenamefont {de~Nova}\ \emph {et~al.}(2016)\citenamefont
  {de~Nova}, \citenamefont {Finazzi},\ and\ \citenamefont
  {Carusotto}}]{Nova_1}%
  \BibitemOpen
  \bibfield  {author} {\bibinfo {author} {\bibfnamefont {J.~R.~M.}\
  \bibnamefont {de~Nova}}, \bibinfo {author} {\bibfnamefont {S.}~\bibnamefont
  {Finazzi}}, \ and\ \bibinfo {author} {\bibfnamefont {I.}~\bibnamefont
  {Carusotto}},\ }\bibfield  {title} {\enquote {\bibinfo {title}
  {Time-dependent study of a black-hole laser in a flowing atomic
  condensate},}\ }\href {\doibase 10.1103/PhysRevA.94.043616} {\bibfield
  {journal} {\bibinfo  {journal} {Phys. Rev. A}\ }\textbf {\bibinfo {volume}
  {94}},\ \bibinfo {pages} {043616} (\bibinfo {year} {2016})}\BibitemShut
  {NoStop}%
\bibitem [{\citenamefont {Pavloff}(2002)}]{Pavloff_1}%
  \BibitemOpen
  \bibfield  {author} {\bibinfo {author} {\bibfnamefont {Nicolas}\ \bibnamefont
  {Pavloff}},\ }\bibfield  {title} {\enquote {\bibinfo {title} {Breakdown of
  superfluidity of an atom laser past an obstacle},}\ }\href {\doibase
  10.1103/PhysRevA.66.013610} {\bibfield  {journal} {\bibinfo  {journal} {Phys.
  Rev. A}\ }\textbf {\bibinfo {volume} {66}},\ \bibinfo {pages} {013610}
  (\bibinfo {year} {2002})}\BibitemShut {NoStop}%
\bibitem [{\citenamefont {Abdullaev}\ \emph {et~al.}(2012)\citenamefont
  {Abdullaev}, \citenamefont {Galimzyanov},\ and\ \citenamefont
  {Ismatullaev}}]{Abdullaev_1}%
  \BibitemOpen
  \bibfield  {author} {\bibinfo {author} {\bibfnamefont {F.Kh.}\ \bibnamefont
  {Abdullaev}}, \bibinfo {author} {\bibfnamefont {R.M.}\ \bibnamefont
  {Galimzyanov}}, \ and\ \bibinfo {author} {\bibfnamefont {Kh.N.}\ \bibnamefont
  {Ismatullaev}},\ }\bibfield  {title} {\enquote {\bibinfo {title} {Quasi 1d
  bose–einstein condensate flow past a nonlinear barrier},}\ }\href@noop {}
  {\bibfield  {journal} {\bibinfo  {journal} {Physics Letters A}\ }\textbf
  {\bibinfo {volume} {376}},\ \bibinfo {pages} {3372 -- 3376} (\bibinfo {year}
  {2012})}\BibitemShut {NoStop}%
\bibitem [{\citenamefont {Frisch}\ \emph {et~al.}(1992)\citenamefont {Frisch},
  \citenamefont {Pomeau},\ and\ \citenamefont {Rica}}]{Frisch_1}%
  \BibitemOpen
  \bibfield  {author} {\bibinfo {author} {\bibfnamefont {T.}~\bibnamefont
  {Frisch}}, \bibinfo {author} {\bibfnamefont {Y.}~\bibnamefont {Pomeau}}, \
  and\ \bibinfo {author} {\bibfnamefont {S.}~\bibnamefont {Rica}},\ }\bibfield
  {title} {\enquote {\bibinfo {title} {Transition to dissipation in a model of
  superflow},}\ }\href {\doibase 10.1103/PhysRevLett.69.1644} {\bibfield
  {journal} {\bibinfo  {journal} {Phys. Rev. Lett.}\ }\textbf {\bibinfo
  {volume} {69}},\ \bibinfo {pages} {1644--1647} (\bibinfo {year}
  {1992})}\BibitemShut {NoStop}%
\bibitem [{\citenamefont {Engels}\ and\ \citenamefont
  {Atherton}(2007)}]{Engels_1}%
  \BibitemOpen
  \bibfield  {author} {\bibinfo {author} {\bibfnamefont {P.}~\bibnamefont
  {Engels}}\ and\ \bibinfo {author} {\bibfnamefont {C.}~\bibnamefont
  {Atherton}},\ }\bibfield  {title} {\enquote {\bibinfo {title} {Stationary and
  nonstationary fluid flow of a bose-einstein condensate through a penetrable
  barrier},}\ }\href {\doibase 10.1103/PhysRevLett.99.160405} {\bibfield
  {journal} {\bibinfo  {journal} {Phys. Rev. Lett.}\ }\textbf {\bibinfo
  {volume} {99}},\ \bibinfo {pages} {160405} (\bibinfo {year}
  {2007})}\BibitemShut {NoStop}%
\bibitem [{\citenamefont {Ee}\ \emph {et~al.}(2011)\citenamefont {Ee},
  \citenamefont {Grimshaw}, \citenamefont {Chow},\ and\ \citenamefont
  {Zhang}}]{Bernard_1}%
  \BibitemOpen
  \bibfield  {author} {\bibinfo {author} {\bibfnamefont {Bernard~K.}\
  \bibnamefont {Ee}}, \bibinfo {author} {\bibfnamefont {R.~H.~J.}\ \bibnamefont
  {Grimshaw}}, \bibinfo {author} {\bibfnamefont {K.~W.}\ \bibnamefont {Chow}},
  \ and\ \bibinfo {author} {\bibfnamefont {D-H.}\ \bibnamefont {Zhang}},\
  }\bibfield  {title} {\enquote {\bibinfo {title} {Steady transcritical flow
  over an obstacle: Parametric map of solutions of the forced extended
  korteweg–de vries equation},}\ }\href {\doibase 10.1063/1.3582523}
  {\bibfield  {journal} {\bibinfo  {journal} {Physics of Fluids}\ }\textbf
  {\bibinfo {volume} {23}},\ \bibinfo {pages} {046602} (\bibinfo {year}
  {2011})},\ \Eprint {http://arxiv.org/abs/http://dx.doi.org/10.1063/1.3582523}
  {http://dx.doi.org/10.1063/1.3582523} \BibitemShut {NoStop}%
\bibitem [{\citenamefont {Leszczyszyn}\ \emph {et~al.}(2009)\citenamefont
  {Leszczyszyn}, \citenamefont {El}, \citenamefont {Gladush},\ and\
  \citenamefont {Kamchatnov}}]{Leszczyszyn_1}%
  \BibitemOpen
  \bibfield  {author} {\bibinfo {author} {\bibfnamefont {A.~M.}\ \bibnamefont
  {Leszczyszyn}}, \bibinfo {author} {\bibfnamefont {G.~A.}\ \bibnamefont {El}},
  \bibinfo {author} {\bibfnamefont {Yu.~G.}\ \bibnamefont {Gladush}}, \ and\
  \bibinfo {author} {\bibfnamefont {A.~M.}\ \bibnamefont {Kamchatnov}},\
  }\bibfield  {title} {\enquote {\bibinfo {title} {Transcritical flow of a
  bose-einstein condensate through a penetrable barrier},}\ }\href {\doibase
  10.1103/PhysRevA.79.063608} {\bibfield  {journal} {\bibinfo  {journal} {Phys.
  Rev. A}\ }\textbf {\bibinfo {volume} {79}},\ \bibinfo {pages} {063608}
  (\bibinfo {year} {2009})}\BibitemShut {NoStop}%
\bibitem [{\citenamefont {Recati}\ \emph {et~al.}(2009)\citenamefont {Recati},
  \citenamefont {Pavloff},\ and\ \citenamefont {Carusotto}}]{Recati_1}%
  \BibitemOpen
  \bibfield  {author} {\bibinfo {author} {\bibfnamefont {A.}~\bibnamefont
  {Recati}}, \bibinfo {author} {\bibfnamefont {N.}~\bibnamefont {Pavloff}}, \
  and\ \bibinfo {author} {\bibfnamefont {I.}~\bibnamefont {Carusotto}},\
  }\bibfield  {title} {\enquote {\bibinfo {title} {Bogoliubov theory of
  acoustic hawking radiation in bose-einstein condensates},}\ }\href {\doibase
  10.1103/PhysRevA.80.043603} {\bibfield  {journal} {\bibinfo  {journal} {Phys.
  Rev. A}\ }\textbf {\bibinfo {volume} {80}},\ \bibinfo {pages} {043603}
  (\bibinfo {year} {2009})}\BibitemShut {NoStop}%
\bibitem [{\citenamefont {Finazzi}\ \emph {et~al.}(2015)\citenamefont
  {Finazzi}, \citenamefont {Piazza}, \citenamefont {Abad}, \citenamefont
  {Smerzi},\ and\ \citenamefont {Recati}}]{Finazzi_1}%
  \BibitemOpen
  \bibfield  {author} {\bibinfo {author} {\bibfnamefont {S.}~\bibnamefont
  {Finazzi}}, \bibinfo {author} {\bibfnamefont {F.}~\bibnamefont {Piazza}},
  \bibinfo {author} {\bibfnamefont {M.}~\bibnamefont {Abad}}, \bibinfo {author}
  {\bibfnamefont {A.}~\bibnamefont {Smerzi}}, \ and\ \bibinfo {author}
  {\bibfnamefont {A.}~\bibnamefont {Recati}},\ }\bibfield  {title} {\enquote
  {\bibinfo {title} {Instability of the superfluid flow as black-hole lasing
  effect},}\ }\href {\doibase 10.1103/PhysRevLett.114.245301} {\bibfield
  {journal} {\bibinfo  {journal} {Phys. Rev. Lett.}\ }\textbf {\bibinfo
  {volume} {114}},\ \bibinfo {pages} {245301} (\bibinfo {year}
  {2015})}\BibitemShut {NoStop}%
\bibitem [{\citenamefont {Bobrovska}\ \emph {et~al.}(2014)\citenamefont
  {Bobrovska}, \citenamefont {Ostrovskaya},\ and\ \citenamefont
  {Matuszewski}}]{Natalia_1}%
  \BibitemOpen
  \bibfield  {author} {\bibinfo {author} {\bibfnamefont {Nataliya}\
  \bibnamefont {Bobrovska}}, \bibinfo {author} {\bibfnamefont {Elena~A.}\
  \bibnamefont {Ostrovskaya}}, \ and\ \bibinfo {author} {\bibfnamefont
  {Micha\l{}}\ \bibnamefont {Matuszewski}},\ }\bibfield  {title} {\enquote
  {\bibinfo {title} {Stability and spatial coherence of nonresonantly pumped
  exciton-polariton condensates},}\ }\href {\doibase
  10.1103/PhysRevB.90.205304} {\bibfield  {journal} {\bibinfo  {journal} {Phys.
  Rev. B}\ }\textbf {\bibinfo {volume} {90}},\ \bibinfo {pages} {205304}
  (\bibinfo {year} {2014})}\BibitemShut {NoStop}%
\bibitem [{\citenamefont {Pham}\ and\ \citenamefont {Brachet}(2002)}]{Pham_1}%
  \BibitemOpen
  \bibfield  {author} {\bibinfo {author} {\bibfnamefont {Chi-Tuong}\
  \bibnamefont {Pham}}\ and\ \bibinfo {author} {\bibfnamefont {Marc}\
  \bibnamefont {Brachet}},\ }\bibfield  {title} {\enquote {\bibinfo {title}
  {Dynamical scaling laws in two types of extended hamiltonian systems at
  dissipation onset},}\ }\href {\doibase
  https://doi.org/10.1016/S0167-2789(01)00388-8} {\bibfield  {journal}
  {\bibinfo  {journal} {Physica D: Nonlinear Phenomena}\ }\textbf {\bibinfo
  {volume} {163}},\ \bibinfo {pages} {127 -- 149} (\bibinfo {year}
  {2002})}\BibitemShut {NoStop}%
\bibitem [{\citenamefont {Wouters}\ and\ \citenamefont
  {Carusotto}(2007)}]{Wouters_1}%
  \BibitemOpen
  \bibfield  {author} {\bibinfo {author} {\bibfnamefont {Michiel}\ \bibnamefont
  {Wouters}}\ and\ \bibinfo {author} {\bibfnamefont {Iacopo}\ \bibnamefont
  {Carusotto}},\ }\bibfield  {title} {\enquote {\bibinfo {title} {Excitations
  in a nonequilibrium bose-einstein condensate of exciton polaritons},}\ }\href
  {\doibase 10.1103/PhysRevLett.99.140402} {\bibfield  {journal} {\bibinfo
  {journal} {Phys. Rev. Lett.}\ }\textbf {\bibinfo {volume} {99}},\ \bibinfo
  {pages} {140402} (\bibinfo {year} {2007})}\BibitemShut {NoStop}%
\bibitem [{\citenamefont {Theocharis}\ \emph {et~al.}(2005)\citenamefont
  {Theocharis}, \citenamefont {Schmelcher}, \citenamefont {Oberthaler},
  \citenamefont {Kevrekidis},\ and\ \citenamefont
  {Frantzeskakis}}]{Theocharis_1}%
  \BibitemOpen
  \bibfield  {author} {\bibinfo {author} {\bibfnamefont {G.}~\bibnamefont
  {Theocharis}}, \bibinfo {author} {\bibfnamefont {P.}~\bibnamefont
  {Schmelcher}}, \bibinfo {author} {\bibfnamefont {M.~K.}\ \bibnamefont
  {Oberthaler}}, \bibinfo {author} {\bibfnamefont {P.~G.}\ \bibnamefont
  {Kevrekidis}}, \ and\ \bibinfo {author} {\bibfnamefont {D.~J.}\ \bibnamefont
  {Frantzeskakis}},\ }\bibfield  {title} {\enquote {\bibinfo {title}
  {Lagrangian approach to the dynamics of dark matter-wave solitons},}\ }\href
  {\doibase 10.1103/PhysRevA.72.023609} {\bibfield  {journal} {\bibinfo
  {journal} {Phys. Rev. A}\ }\textbf {\bibinfo {volume} {72}},\ \bibinfo
  {pages} {023609} (\bibinfo {year} {2005})}\BibitemShut {NoStop}%
\bibitem [{\citenamefont {Amo}\ \emph {et~al.}(2011)\citenamefont {Amo},
  \citenamefont {Pigeon}, \citenamefont {Sanvitto}, \citenamefont {Sala},
  \citenamefont {Hivet}, \citenamefont {Carusotto}, \citenamefont {Pisanello},
  \citenamefont {Lem{\'e}nager}, \citenamefont {Houdr{\'e}}, \citenamefont
  {Giacobino}, \citenamefont {Ciuti},\ and\ \citenamefont {Bramati}}]{Amo_2}%
  \BibitemOpen
  \bibfield  {author} {\bibinfo {author} {\bibfnamefont {A.}~\bibnamefont
  {Amo}}, \bibinfo {author} {\bibfnamefont {S.}~\bibnamefont {Pigeon}},
  \bibinfo {author} {\bibfnamefont {D.}~\bibnamefont {Sanvitto}}, \bibinfo
  {author} {\bibfnamefont {V.~G.}\ \bibnamefont {Sala}}, \bibinfo {author}
  {\bibfnamefont {R.}~\bibnamefont {Hivet}}, \bibinfo {author} {\bibfnamefont
  {I.}~\bibnamefont {Carusotto}}, \bibinfo {author} {\bibfnamefont
  {F.}~\bibnamefont {Pisanello}}, \bibinfo {author} {\bibfnamefont
  {G.}~\bibnamefont {Lem{\'e}nager}}, \bibinfo {author} {\bibfnamefont
  {R.}~\bibnamefont {Houdr{\'e}}}, \bibinfo {author} {\bibfnamefont
  {E}~\bibnamefont {Giacobino}}, \bibinfo {author} {\bibfnamefont
  {C.}~\bibnamefont {Ciuti}}, \ and\ \bibinfo {author} {\bibfnamefont
  {A.}~\bibnamefont {Bramati}},\ }\bibfield  {title} {\enquote {\bibinfo
  {title} {Polariton superfluids reveal quantum hydrodynamic solitons},}\
  }\href {\doibase 10.1126/science.1202307} {\bibfield  {journal} {\bibinfo
  {journal} {Science}\ }\textbf {\bibinfo {volume} {332}},\ \bibinfo {pages}
  {1167--1170} (\bibinfo {year} {2011})},\ \Eprint
  {http://arxiv.org/abs/http://science.sciencemag.org/content/332/6034/1167.full.pdf}
  {http://science.sciencemag.org/content/332/6034/1167.full.pdf} \BibitemShut
  {NoStop}%
\bibitem [{\citenamefont {Van~Regemortel}\ and\ \citenamefont
  {Wouters}(2014)}]{Regemortel_1}%
  \BibitemOpen
  \bibfield  {author} {\bibinfo {author} {\bibfnamefont {Mathias}\ \bibnamefont
  {Van~Regemortel}}\ and\ \bibinfo {author} {\bibfnamefont {Michiel}\
  \bibnamefont {Wouters}},\ }\bibfield  {title} {\enquote {\bibinfo {title}
  {Negative drag in nonequilibrium polariton quantum fluids},}\ }\href
  {\doibase 10.1103/PhysRevB.89.085303} {\bibfield  {journal} {\bibinfo
  {journal} {Phys. Rev. B}\ }\textbf {\bibinfo {volume} {89}},\ \bibinfo
  {pages} {085303} (\bibinfo {year} {2014})}\BibitemShut {NoStop}%
\bibitem [{\citenamefont {Bobrovska}\ and\ \citenamefont
  {Matuszewski}(2015)}]{Natalia_2}%
  \BibitemOpen
  \bibfield  {author} {\bibinfo {author} {\bibfnamefont {Nataliya}\
  \bibnamefont {Bobrovska}}\ and\ \bibinfo {author} {\bibfnamefont {Micha\l{}}\
  \bibnamefont {Matuszewski}},\ }\bibfield  {title} {\enquote {\bibinfo {title}
  {Adiabatic approximation and fluctuations in exciton-polariton
  condensates},}\ }\href {\doibase 10.1103/PhysRevB.92.035311} {\bibfield
  {journal} {\bibinfo  {journal} {Phys. Rev. B}\ }\textbf {\bibinfo {volume}
  {92}},\ \bibinfo {pages} {035311} (\bibinfo {year} {2015})}\BibitemShut
  {NoStop}%
\bibitem [{\citenamefont {Stiller}\ \emph {et~al.}(1995)\citenamefont
  {Stiller}, \citenamefont {Popp},\ and\ \citenamefont
  {Kramer}}]{Popp_FromDarkSolitons}%
  \BibitemOpen
  \bibfield  {author} {\bibinfo {author} {\bibfnamefont {Olaf}\ \bibnamefont
  {Stiller}}, \bibinfo {author} {\bibfnamefont {Stefan}\ \bibnamefont {Popp}},
  \ and\ \bibinfo {author} {\bibfnamefont {Lorenz}\ \bibnamefont {Kramer}},\
  }\bibfield  {title} {\enquote {\bibinfo {title} {From dark solitons in the
  defocusing nonlinear schro¨dinger to holes in the complex ginzburg-landau
  equation},}\ }\href {\doibase https://doi.org/10.1016/0167-2789(95)00071-B}
  {\bibfield  {journal} {\bibinfo  {journal} {Physica D: Nonlinear Phenomena}\
  }\textbf {\bibinfo {volume} {84}},\ \bibinfo {pages} {424 -- 436} (\bibinfo
  {year} {1995})}\BibitemShut {NoStop}%
\bibitem [{\citenamefont {Bekki}\ and\ \citenamefont
  {Nozaki}(1985)}]{Bekki_Holes}%
  \BibitemOpen
  \bibfield  {author} {\bibinfo {author} {\bibfnamefont {Naoaki}\ \bibnamefont
  {Bekki}}\ and\ \bibinfo {author} {\bibfnamefont {Kazuhiro}\ \bibnamefont
  {Nozaki}},\ }\bibfield  {title} {\enquote {\bibinfo {title} {Formations of
  spatial patterns and holes in the generalized ginzburg-landau equation},}\
  }\href {\doibase https://doi.org/10.1016/0375-9601(85)90759-5} {\bibfield
  {journal} {\bibinfo  {journal} {Physics Letters A}\ }\textbf {\bibinfo
  {volume} {110}},\ \bibinfo {pages} {133 -- 135} (\bibinfo {year}
  {1985})}\BibitemShut {NoStop}%
\bibitem [{\citenamefont {Popp}\ \emph {et~al.}(1993)\citenamefont {Popp},
  \citenamefont {Stiller}, \citenamefont {Aranson}, \citenamefont {Weber},\
  and\ \citenamefont {Kramer}}]{Popp_HoleStability}%
  \BibitemOpen
  \bibfield  {author} {\bibinfo {author} {\bibfnamefont {Stefan}\ \bibnamefont
  {Popp}}, \bibinfo {author} {\bibfnamefont {Olaf}\ \bibnamefont {Stiller}},
  \bibinfo {author} {\bibfnamefont {Igor}\ \bibnamefont {Aranson}}, \bibinfo
  {author} {\bibfnamefont {Andreas}\ \bibnamefont {Weber}}, \ and\ \bibinfo
  {author} {\bibfnamefont {Lorenz}\ \bibnamefont {Kramer}},\ }\bibfield
  {title} {\enquote {\bibinfo {title} {Localized hole solutions and
  spatiotemporal chaos in the 1d complex ginzburg-landau equation},}\ }\href
  {\doibase 10.1103/PhysRevLett.70.3880} {\bibfield  {journal} {\bibinfo
  {journal} {Phys. Rev. Lett.}\ }\textbf {\bibinfo {volume} {70}},\ \bibinfo
  {pages} {3880--3883} (\bibinfo {year} {1993})}\BibitemShut {NoStop}%
\end{thebibliography}%
\end{document}